\documentclass[fleqn,usenatbib]{mnras}

\usepackage{newtxtext,newtxmath}

\usepackage[T1]{fontenc}
\usepackage{bm}
\usepackage{booktabs}
\usepackage{array}
\usepackage{enumerate}
\usepackage{ulem}
\usepackage{threeparttable}
\usepackage{multirow}
\usepackage{soul}
\setstcolor{red}
\usepackage{color}
\newcommand{\add}[1]{#1}

\DeclareRobustCommand{\VAN}[3]{#2}
\let\VANthebibliography\thebibliography
\def\thebibliography{\DeclareRobustCommand{\VAN}[3]{##3}\VANthebibliography}

\usepackage{graphicx}	
\usepackage{amsmath}	

\title[Neutron star--white dwarf Mergers]{Prospects for detecting neutron
star--white dwarf mergers with decihertz gravitational-wave observatories}

\author[Y. Kang, et al.]{Yacheng Kang,$^{1,2}$
Chang Liu,$^{1,2,3}$
Jin-Ping Zhu,$^{4,5}$
Yong Gao,$^{1,2,6}$
Lijing Shao,$^{2,7}$\thanks{E-mail: lshao@pku.edu.cn}
Bing Zhang,$^{8,9}$\thanks{E-mail: bing.zhang@unlv.edu}\newauthor
Hui Sun,$^{7}$
Yi-Han Iris Yin$^{10},$
and Bin-Bin Zhang$^{10,11,12}$
\\
$^{1}$Department of Astronomy, School of Physics, Peking University, Beijing 100871, China\\
$^{2}$Kavli Institute for Astronomy and Astrophysics, Peking University, Beijing 100871, China\\
$^{3}$Laboratoire des 2 Infinis - Toulouse (L2IT-IN2P3), Universit\'e de Toulouse, CNRS, UPS, F-31062 Toulouse Cedex 9, France\\
$^{4}$School of Physics and Astronomy, Monash University, Clayton Victoria 3800, Australia\\
$^{5}$Monash Centre for Astrophysics, School of Physics and Astronomy, Monash University, Clayton, VIC 3800, Australia\\
$^{6}$Max Planck Institute for Gravitational Physics (Albert Einstein Institute), Am M\"uhlenberg 1, Potsdam 14476, Germany\\
$^{7}$National Astronomical Observatories, Chinese Academy of Sciences, Beijing 100012, China\\
$^{8}$Nevada Center for Astrophysics, University of Nevada, Las Vegas, NV 89154, USA\\
$^{9}$Department of Physics and Astronomy, University of Nevada, Las Vegas, NV 89154, USA\\
$^{10}$School of Astronomy and Space Science, Nanjing University, Nanjing 210093, China\\
$^{11}$Key Laboratory of Modern Astronomy and Astrophysics (Nanjing University), Ministry of Education, China\\
$^{12}$Purple Mountain Observatory, Chinese Academy of Sciences, Nanjing 210023, China}

\date{Accepted XXX. Received YYY; in original form ZZZ}

\pubyear{2023}

\begin{document}
\label{firstpage}
\pagerange{\pageref{firstpage}--\pageref{lastpage}}
\maketitle

\begin{abstract}
Based on different neutron star--white dwarf (NS-WD) population models, we
investigate the prospects of gravitational-wave (GW) detections for NS-WD
mergers, with the help of early warnings from two space-borne decihertz GW
observatories, DO-Optimal and DECIGO. We not only give quick assessments of the
GW detection rates for NS-WD mergers with the two decihertz GW detectors, but
also report systematic analyses on the characteristics of GW-detectable merger
events using the method of Fisher matrix. With a sufficient one-day
early-warning time, the yearly GW detection number for DO-Optimal is in the
range of $ (1.5$--$1.9) \times 10^{3}$, while it is $ (3.3$--$4.6) \times
10^{4}$  for DECIGO. More importantly, our results show that most NS-WD mergers
can be localized with an uncertainty of $\mathcal{O}(10^{-2})\,\mathrm{deg}^2$.
Given the NS-WD merger as a possible origin for a peculiar long-duration gamma-ray
burst, GRB\,211211A, \add{followed with kilonova-like emissions,} we further suggest that the GW early-warning
detection would allow future electromagnetic telescopes to get prepared to
follow-up transients after some special NS-WD mergers. Based on our analyses, we
emphasize that such a feasible ``wait-for'' pattern can help to firmly identify
the origin of GRB\,211211A-like events in the future and bring excellent
opportunities for the multimessenger astronomy.
\end{abstract}




\section{Introduction}
\label{ sec:intro }

The mergers of double-compact-object (DCO) systems have been explored
extensively, both as progenitors of various electromagnetic (EM) transients and as
gravitational wave (GW) sources. For example, binary white dwarfs (BWDs) are not
only one of the most promising candidates of type Ia supernova (SN) progenitors
\citep{Livio:2018rue, Soker:2017cbn, Wang:2018pac, Ruiter:2019tam}, but are also
expected to be the dominant GW sources by numbers in our Milky Way (MW) for the
Laser Interferometer Space Antenna (LISA) mission\footnote{LISA is an ESA-led
space-borne GW observatory, with NASA as a junior partner. It is due to be
launched in 2034 with the aim to record and study GWs in the millihertz
frequency band. More information about the LISA mission can be found at 
\url{https://lisa.nasa.gov}.} \citep{Lamberts:2018cge, Bayle:2022hvs,
LISA:2022yao}. For compact binary mergers containing at least one neutron star
(NS), it has long been proposed that the mergers of binary neutron stars (BNSs)
and neutron star--black holes (NS-BHs) are the progenitors of short-duration
gamma-ray bursts \citep[sGRBs; ][]{Paczynski:1986px, Eichler:1989ve,
Narayan:1992iy, Zhang:2018ads}. The milestone co-detection of GW and
EM signals during GW170817/GRB\,170817A/AT\,2017gfo  events
\citep{LIGOScientific:2017vwq, LIGOScientific:2017zic, Goldstein:2017mmi,
Savchenko:2017ffs, Coulter:2017wya, Evans:2017mmy,
Pian:2017gtc, Kilpatrick:2017mhz, 2018NatCo...9..447Z} has not only provided us a smoking-gun
evidence for the BNS merger origin of sGRBs and kilonovae \citep{Li:1998bw}, but
also an unprecedented opportunity to explore gravity theories and fundamental
physics in extreme environments \citep{LIGOScientific:2017pwl,
LIGOScientific:2018cki, LIGOScientific:2018dkp, Shao:2017gwu,
Sathyaprakash:2019yqt, Arun:2022vqj}.

Nevertheless, the mergers of other kinds of DCO systems, such as NS-WD binaries,
have not been explored to the same depth in literature. Until recently,
\citet{Yang:2022qmy} adopted a NS-WD merger scenario to explain the
kilonova-like emission following a peculiar long-duration gamma-ray burst
(lGRB), GRB\,211211A \citep{Rastinejad:2022zbg, Troja:2022yya, Mei:2022ncd,
Gompertz:2022jsg}, attracting significant attention on NS-WD mergers and their
detection prospects \citep{Zhong:2023zwh, Yin:2023gwc}. Compared with other
types of DCO systems (e.g., BWDs, BNSs, NS-BHs, {\it etc.}), the different
component properties of NS-WD binaries could have significant impacts on the
early-time dynamics of mass-transfer phase, initial conditions of accretion
disk, and final merger remnants, {\it etc.} \citep{King:2006nw,
Chattopadhyay:2007rx, Paschalidis:2011ez, Margalit:2016joe, Margalit:2016hlx}.
In particular, \citet{Zhong:2023zwh} showed that GRB\,211211A-like events could
arise from a NS-WD merger if the central engine leaves a magnetar behind. They
proposed that the magnetic bubble eruptions from the toroidal magnetic field
amplification of the pre-merger NS could successfully produce the main burst of
GRB\,211211A.  NS-WD systems are likely the most common type of DCO systems
besides BWDs \citep{Nelemans:2001hp, Toonen:2018njy}, and may have various
observable explosive transients \citep{Metzger:2011zk, Zenati:2018gcp,
Fernandez:2019eqo, Gillanders:2020fhm, Bobrick:2021hho, Kaltenborn:2022vxf},
which will bring excellent opportunities for future multimessenger astronomy.
Overall, it is worth analyzing the consequences of NS-WD mergers in more detail.

Based on the above arguments, in this work we discuss the realistic prospects of
detecting NS-WD mergers with GW early warnings using space-borne decihertz GW
detectors. Given the lower sensitive frequency range ($0.01$--$1\,\mathrm{Hz}$)
of decihertz detectors, they can offer alerts to NS-WD mergers much earlier than
the current and future ground-based detectors\footnote{\add{Ground-based GW detectors include the current Advanced Laser Interferometer Gravitational-wave Observatory (LIGO), the Advanced Virgo, the Kamioka Gravitational Wave Detector (KAGRA), and the future next-generation detectors, such as the Einstein Telescope and the Cosmic Explorer \citep{KAGRA:2013rdx, Reitze:2019iox, Maggiore:2019uih, Ronchini:2022gwk, Banerjee:2022gkv, Branchesi:2023mws}.}} (usually sensitive in the $10$--$10^4\,\mathrm{Hz}$
band), or even earlier than the EM facilities \citep{Liu:2022mcd, Kang:2022nmz, 2023SSPMA..53j0014K}.
In addition, another motivation to consider decihertz GW early warnings is that
the upper cutoff GW frequency is expected to be $\lesssim 1$\,Hz for NS-WD
mergers in their inspiral phase. This is because the runaway mass-transfer phase
and tidal disruption outcome occur earlier for NS-WD mergers
\citep{1988ApJ...332..193V, Paschalidis:2009zz, Margalit:2016joe,
Kaltenborn:2022vxf}, especially when compared with BNS and NS-BH merger
events.\footnote{Note that for BNS and NS-BH mergers in the inspiral phase, the
upper cutoff GW frequency can be approximated by the GW frequency at the
innermost stable circular orbit.} In view of this, it would be difficult for the
NS-WD inspiral GW signals to enter the ground-based GW detectors, unlike the
situation for more massive DCO systems. Nevertheless, it allows us to propose a
feasible ``wait-for'' GW detecting mode for NS-WD mergers with decihertz GW
detectors, and provide helpful inputs for future multimessenger astrophysics.

Before quantitatively analysing the early warnings from two representative
decihertz GW observatories, DO-Optimal \citep[DO-OPT;][]{Sedda:2019uro,
Sedda:2021yhn} and DECIGO \citep[DEC;][]{Kawamura:2011zz, Kawamura:2020pcg}, we
first use the method of convolution with different star formation rates (SFRs)
and delay-time distributions (DTDs) to obtain four kinds of NS-WD merger
population models. Given the short tidal-disruption timescale \citep[see,
e.g.,][]{Kaltenborn:2022vxf},  $\lesssim {\cal O}(\mathrm{min})$,  and possibly
observable EM transients \citep{King:2006nw, Chattopadhyay:2007rx,
Metzger:2011zk, Margalit:2016joe, Margalit:2016hlx, Yang:2022qmy} for NS-WD
mergers, we follow \citet{Kang:2022nmz} and set the early-warning time to be
$t_\mathrm{e} = 1\,\mathrm{d}$ and signal-to-noise ratio (SNR) threshold value
to be 8 for our GW detection strategy. Differently from BNS mergers or NS-BH
mergers, we define the time when the WD starts to experience the Roche-lobe
overflow as the merger time for NS-WD binaries, considering that there is no
well-modeled GW waveform during the runaway mass-transfer phase. With the aim to
extend early studies on decihertz GW alerts for NS-WD mergers, we not only give
quick assessments of yearly detection numbers and percentages with DO-OPT and
DEC, but also report more detailed analyses on the characteristics of
GW-detectable merger events using the Fisher matrix. We find that DEC has better
performance than DO-OPT as a whole on GW early-warning detections and
localization abilities, especially for high-redshift (high-$z$) events. For
those mergers that would yield the best estimation results of distance and
angular resolution, we collect them into a Golden Sample Set, and present the
detection rates of NS-WD mergers in the Golden Sample Set for different
population models with DO-OPT and DEC.

Taking the recent peculiar lGRB, GRB\,211211A, as an example, we further suggest
that the GW early-warning detection would allow future EM telescopes to get
prepared for possible follow-up transients after some special NS-WD mergers.
Although there are a few groups suggesting different origins for
GRB\,211211A,\footnote{For examples, \citet{Waxman:2022pfm} have shown that the
thermal emission from dust could explain the observed near-infrared data;
\citet{Zhu:2022kbt} have suggested that a NS-BH merger could roughly reproduce
the multiwavelength observations; \citet{Gompertz:2022jsg} concluded that the
spectral evolution could be explained by a transition from a fast-cooling mode
to a slow-cooling regime, favoring a BNS merger scenario rather than a NS-BH
origin; \citet{Barnes:2023ixp} also found that the afterglow-subtracted emission
of GRB\,211211A is in best agreement for collapsar models with high kinetic
energies. Unfortunately, GRB\,211211A was detected prior to the fourth observing
run of the LIGO-Virgo-KAGRA Collaboration. Overall, the origin of GRB\,211211A
is still under debate.} at least most studies have supported or directly
considered compact star merger scenarios for this event, namely compact-binary
lGRBs (cb-lGRBs), given the features of kilonova-like emissions and host-galaxy
properties including the offset, {\it etc.}. Regardless of the exact composition
of this binary system, with a sufficient early-warning time ($t_\mathrm{e} =
1\,\mathrm{d}$) and localization accuracies ($\Delta \Omega \lesssim
1\,\mathrm{deg}^2$), we point out that one can prepare well in advance for
future EM transients of GRB\,211211A-like events, and no longer needs to
consider the field of view (FoV) discounts and complex searching
strategies.\footnote{We refer readers to \citet{Kang:2022nmz} for more 
descriptions.} Such a feasible wait-for pattern, if realized in future, can help
to firmly identify the origin of GRB\,211211A-like events and enhance the
comprehension of GRB’s physical type.

The organization of this paper is as follows. We first overview the construction
of the NS-WD merger population models in Section~\ref{ sec:populations }. In
Section~\ref{ sec:warnings }, we introduce the GW detecting strategy with two
space-borne decihertz GW observatories. Using the above ingredients, we report
our results and detailed analyses on GW early-warning detections of NS-WD
mergers in Section~\ref{ sec:Results }. Finally, Section~\ref{ sec:Conclusion }
concludes the paper. Throughout this paper, we adopt a standard $\rm \Lambda
CDM$ model with the matter density parameter $\Omega_{\mathrm{m}}=0.315$, the
dark-energy density parameter $\Omega_{\Lambda}=0.685$, and the Hubble-Lemaître
constant $H_{0}=67.4\,\mathrm{km}\,\mathrm{s}^{-1}\,\mathrm{Mpc}^{-1}$
\citep{Planck:2018vyg}.


\begin{table*}
    \renewcommand\arraystretch{2}
    \centering
    \caption{Simulated NS-WD merger numbers per year for different population
    models with $\dot{\rho}_{0} = 390\,\mathrm{Gpc}^{-3}\,\mathrm{yr}^{-1}$. In
    the last row, we list the lower and upper values in brackets, by rescaling
    our results with $\dot{\rho}_{0} \approx [90,
    5800]\,\mathrm{Gpc}^{-3}\,\mathrm{yr}^{-1}$.  More descriptions of the four
    population models are given in Section~\ref{ sec:zdistri }.}
    \setlength{\tabcolsep}{0.75cm}{\begin{tabular}{c c c c}
    \toprule
    \toprule
    \vspace{-3.5em}\\
    \multicolumn{4}{c}{Population Model}\vspace{0.1em}\\  
    \cline{1-4}
    A    & B    & C    & D\\
    \cline{1-4}
      $5.6 \times 10^{5}$    
    & $4.1 \times 10^{5}$    
    & $4.9 \times 10^{5}$    
    & $3.8 \times 10^{5}$\vspace{-1em}\\  
      ($1.3 \times 10^{5}$ --$8.3 \times 10^{6}$)            
    & ($9.5 \times 10^{4}$ -- $6.1 \times 10^{6}$)          
    & ($1.1 \times 10^{5}$ -- $7.4 \times 10^{6}$)            
    & ($8.7 \times 10^{4}$ -- $5.6 \times 10^{6}$)\\
    \bottomrule
    \end{tabular}}
    \label{ tab:Total numbers }
\end{table*}

\section{NS-WD Population} 
\label{ sec:populations }

In this section, we briefly describe how we obtain the NS-WD populations
prepared for later analyses. Following \citet{Sun:2015bda}, we also ignore the
possible redshift evolution of intrinsic system parameters for NS-WD mergers.
Within this framework, one can separately discuss the redshift distribution of
NS-WD systems in Section~\ref{ sec:zdistri }. We show more details about mass
distribution of our NS-WD populations in Section~\ref{ sec:mdistri }.

\subsection{Event rate and redshift distribution}
\label{ sec:zdistri }

The number density per unit time for NS-WD mergers at a given redshift $z$ can
be estimated as,
\begin{equation}
\frac{d \dot{N}}{d z}=\frac{\dot{\rho}_{0} f(z)}{1+z} \frac{d V(z)}{d z} \,,
\label{ eq:ddotNdz }
\end{equation}
where $\dot{\rho}_{0}$ is the local NS-WD merger rate density, $f(z)$ is the
dimensionless redshift distribution factor, and $\frac{d V(z)}{d z}$ is the
comoving volume element, 
\begin{equation}
\frac{d V(z)}{d z}=\frac{c}{H_{0}} \frac{4 \pi D_{\mathrm{L}}^{2}}{(1+z)^{2}
\sqrt{\Omega_{\Lambda}+\Omega_{\mathrm{m}}(1+z)^{3}}} \,,
\label{ eq:dVdz }
\end{equation}
where $c$ is the speed of light, and $D_{\mathrm{L}}$ is the luminosity
distance, 
\begin{equation}
D_{\mathrm{L}}=(1+z) \frac{c}{H_{0}} \int_{0}^{z} \frac{d
z}{\sqrt{\Omega_{\Lambda}+\Omega_{\mathrm{m}}(1+z)^{3}}} \,.
\label{ eq:DL }
\end{equation}

The function $f(z)$ in Equation~(\ref{ eq:ddotNdz }) depends on the DTD of NS-WD
mergers superposed on the SFR. Considering different combinations of SFRs and
DTDs, we illustrate in Appendix~\ref{ sec:appA } the redshift distributions of
$f(z)$ with more details. We consider two kinds of analytical models from
\citet{Yuksel:2008cu} and \citet{Madau:2014bja} for SFRs, abbreviated as `Y08'
and `MD14', respectively; as for DTDs, two models---abbreviated as `$\gamma
\alpha$-$\mathrm{H}$' and `$\gamma \alpha$-$\mathrm{V}$'---are adopted in this
work (see Appendix~\ref{ sec:appA } for more descriptions). The population
models are briefly summarized as follows:
\begin{enumerate}[(1)]
   \item \textbf{Model A}: Y08 + $\gamma \alpha$-$\mathrm{H}$;
   \item \textbf{Model B}: Y08 + $\gamma \alpha$-$\mathrm{V}$;
   \item \textbf{Model C}: MD14 + $\gamma \alpha$-$\mathrm{H}$;
   \item \textbf{Model D}: MD14 + $\gamma \alpha$-$\mathrm{V}$.
\end{enumerate}

For the local NS-WD merger rate density, we adopt $\dot{\rho}_{0} ~ {=
390\,\mathrm{Gpc}^{-3}\,\mathrm{yr}^{-1}}$ from \citet{Zhao:2020jks} as the
fiducial value, which is in agreement with many studies \citep{Nelemans:2000es,
Kim:2004kz, OShaughnessy:2009cxm, Bobrick:2017rlo, Toonen:2018njy}. Moreover, in
a similar way to that in \citet{Liu:2022mcd}, we will rescale our results in
later sections by adjusting $\dot{\rho}_{0}$ in the range of $[8,
500]\,\mathrm{Myr}^{-1}$ per MW-like Galaxy, corresponding to $\dot{\rho}_{0}
\approx [90, 5800]\,\mathrm{Gpc}^{-3}\,\mathrm{yr}^{-1}$ hereafter
\citep{Kaltenborn:2022vxf}.\footnote{In most recent studies, a few authors
obtained the conversion factor with the galactic blue luminosity of $1.7 \times
10^{10}\,\mathrm{~L}_{\mathrm{B}, \odot}$ \citep{Kalogera:2001dz} and the blue
luminosity for the local Universe of $1.98 \times
10^{8}\,\mathrm{~L}_{\mathrm{B}, \odot}\,\mathrm{Mpc}^{-3}$
\citep{Kopparapu:2007ib}, where $\mathrm{~L}_{\mathrm{B}, \odot} = 2.16 \times
10^{33}\,\mathrm{erg}\,\mathrm{s}^{-1}$ is the Solar luminosity in the
$B$-band.} We regard this as a crude but reasonable treatment to assess the
systematic uncertainties in our predictions. The simulated NS-WD merger rates in
the Universe for different population models are listed in Table~\ref{ tab:Total
numbers }.

\subsection{Mass distribution of NS-WD binaries}
\label{ sec:mdistri }

\begin{figure}
    \centering
    \includegraphics[width=8.2cm]{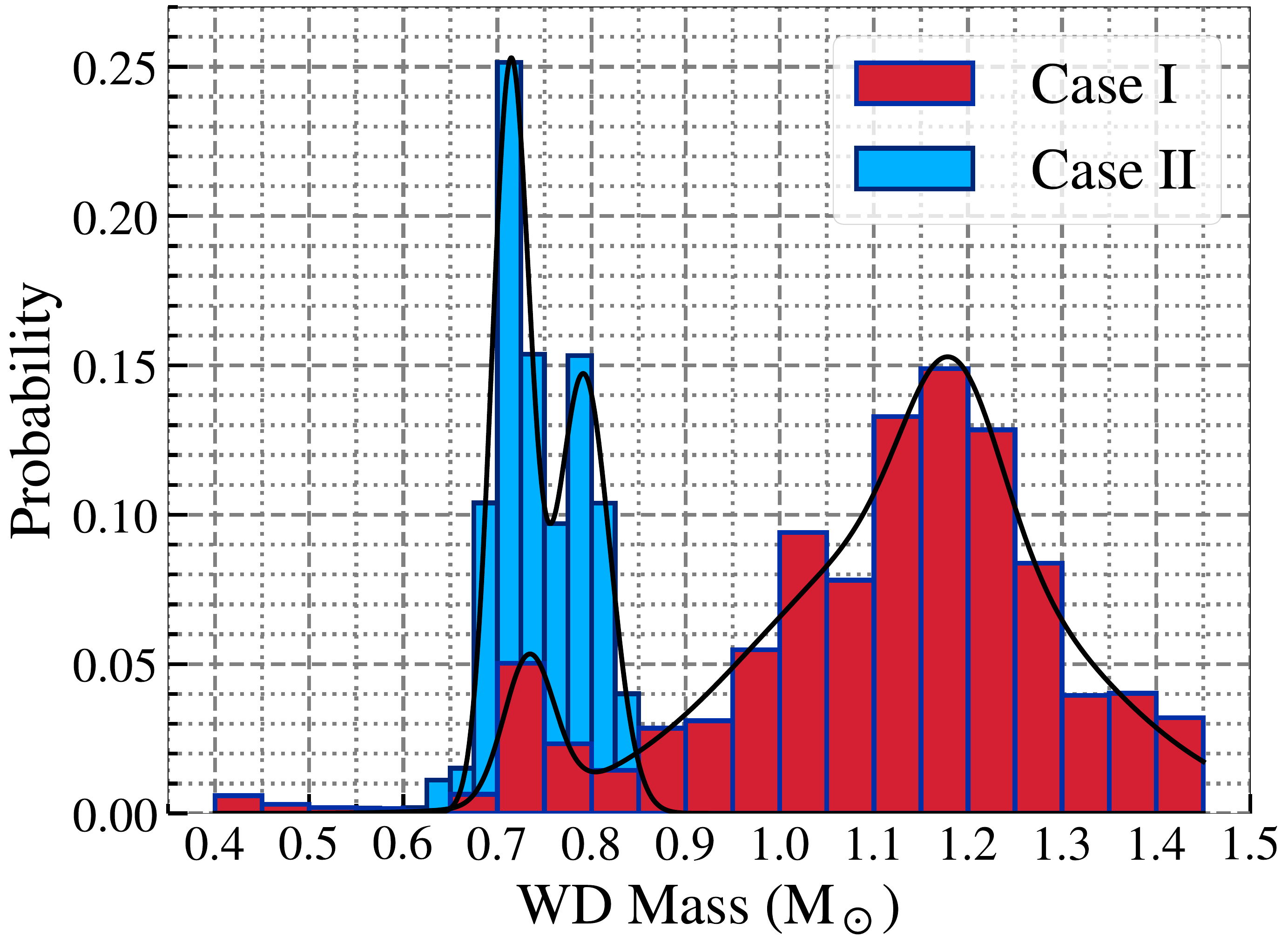}
    \caption{Distributions of the WD component mass obtained by
    \citet{Kaltenborn:2022vxf} for Case I (red) and Case II (blue). With the
    parameters given in Section~\ref{ sec:mdistri }, we also plot the best-fit
    lines (black) for each case. Note that here we have considered
    $M_{\mathrm{WD}} \lesssim M_{\mathrm{ch}}$, where $M_{\mathrm{ch}} \approx
    1.45\,\mathrm{M}_{\odot}$ is the Chandrasekhar mass limit.} 
    \label{ fig:fbm }
\end{figure}

The mass distribution of NS-WD systems is a key ingredient in our calculation of
GW detection rates. However, it is difficult to construct an analytical model
for the probability density function of the component masses in NS-WD binaries.
Therefore, for simplicity, we make use of the population-synthesis simulation
results obtained by \citet{Kaltenborn:2022vxf}. Given that there are two strong
peaks in the mass distribution of NSs at $M_{\mathrm{NS}} =
1.11\,\mathrm{M}_{\odot}$ and $M_{\mathrm{NS}} = 1.26\,\mathrm{M}_{\odot}$
\citep[see Figure~1 in][]{Kaltenborn:2022vxf}, we assume that each NS-WD merger
may have $M_{\mathrm{NS}} = 1.11\,\mathrm{M}_{\odot}$ or $M_{\mathrm{NS}} = ~
{1.26\,\mathrm{M}_{\odot}}$ with an equal probability. The former is referred to
as `Case I', and the latter corresponds to `Case II'. In each case, without
considering the detailed properties of NS-WD systems (e.g., the WD composition,
the DTD models, {\it etc.}),\footnote{This treatment is acceptable because
\citet{Toonen:2018njy} have suggested that there is no significant evolution of
the average WD mass with the delay time for NS-WD binaries. We also find that
the fractions of NS-WD mergers with different WD components (i.e., CO, ONe or He
WDs) remain almost unchanged when different DTD models are adopted \citep[see
Table~2 in][]{Toonen:2018njy}. Moreover, fractions shown in
\citet{Kaltenborn:2022vxf} are also remarkably consistent with the results in
\citet{Toonen:2018njy}.} we choose to fit the mass distribution of WD components
using a model composed of multi-Gaussian components, 
\begin{equation}
\begin{aligned} 
    f^{\mathrm{I}}_\mathrm{M} \propto & \left\{ a_{1} \exp
    \left[-\frac{\left(M_{\mathrm{WD}}-\mu_{1}\right)^{2}}{2
    \sigma_{1}^{2}}\right] +  a_{2} \exp
    \left[-\frac{\left(M_{\mathrm{WD}}-\mu_{2}\right)^{2}}{2
    \sigma_{2}^{2}}\right] \right. \\ 
    & \left. + a_{3} \exp
    \left[-\frac{\left(M_{\mathrm{WD}}-\mu_{3}\right)^{2}}{2
    \sigma_{3}^{2}}\right] \right\} \,,
\end{aligned} 
\label{ eq:mass distri I }
\end{equation}
and 
\begin{equation}
\begin{aligned} 
    f^{\mathrm{II}}_\mathrm{M} \propto & \left\{ a_{4} \exp
    \left[-\frac{\left(M_{\mathrm{WD}}-\mu_{4}\right)^{2}}{2
    \sigma_{4}^{2}}\right] +  a_{5} \exp \left[ -\frac{\left(M_{\mathrm{WD}}
    -\mu_{5}\right)^{2}}{2 \sigma_{5}^{2}} \right] \right\} \,. 
\end{aligned} 
\label{ eq:mass distri II }
\end{equation}
Note that in the above equations, the WD component mass $M_{\mathrm{WD}}$ is in
units of $\mathrm{M}_{\odot}$. We plot the distribution of $M_{\mathrm{WD}}$ for
Case I and Case II in Figure~\ref{ fig:fbm }. The best-fit values of each
parameter in Equation~(\ref{ eq:mass distri I }) and Equation~(\ref{ eq:mass distri II })
are 
\begin{align}
    &a_{1}=0.048\,, \quad
    \mu_{1}=0.73\,, \quad
    \sigma_{1}=0.027\,, \\
    &a_{2}= ~ {0.061} \,, \quad
    \mu_{2}=1.18 \,, \quad
    \sigma_{2}=0.051 \,, \\
    &a_{3}=0.094 \,, \quad 
    \mu_{3}=1.14 \,, \quad 
    \sigma_{3}=0.17 \,,
\end{align}
and
\begin{align}
    &a_{4}=0.25 \,, \quad
    \mu_{4}=0.71 \,, \quad
    \sigma_{4}=0.020 \,, \\
    &a_{5}= {0.15} \,, \quad
    \mu_{5}=0.79 \,, \quad
    \sigma_{5}=0.028 \,.
\end{align}

Finally, let us elaborate on the meaning of `merger' for NS-WD systems in this
work. Regardless of the evolution history, once a NS-WD binary is formed, it
keeps losing its orbital energy through GW emission, causing the orbit to
shrink. When GW emission drives the NS-WD system to the Roche-lobe overflow
(RLOF) orbital separation, the mass from the WD begins to transfer to the NS
companion. A well-known fit of the RLOF separation is given by
\citep{Eggleton:1983rx},
\begin{equation}
a_{\mathrm{RLOF}} \approx R_{\mathrm{WD}} \frac{0.6 q^{2 / 3}+\ln \left(1+q^{1 /
3}\right)}{0.49 q^{2 / 3}}\,,
\label{ eq:RLOF }
\end{equation}
where $q=M_{\mathrm{WD}} / M_{\mathrm{NS}}$ and $R_{\mathrm{WD}}$ is the WD
radius. The latter can be well approximated by \citep{1972ApJ...175..417N},
\begin{equation}
R_{\mathrm{WD}} \approx 10^{9}
\mathrm{~cm}\left(\frac{M_{\mathrm{WD}}}{0.7\,\mathrm{M}_{\odot}}\right)^{-1 /
3}\left[1-\left(\frac{M_{\mathrm{WD}}}{M_{\mathrm{ch}}}\right)^{4 / 3}\right]^{1
/ 2}\,,
\label{ eq:WD radius }
\end{equation}
where $M_{\mathrm{ch}} \approx 1.45\,\mathrm{M}_{\odot}$ is the Chandrasekhar
mass limit assuming the mean molecular weight per electron,
$\mu_{\mathrm{e}}=2$. 

During the mass-transfer phase, the binary separation can increase due to the
conservation of angular momentum; on the other hand, the WD radius increases due
to its mass loss [see Equation~(\ref{ eq:WD radius })], which can increase the
critical RLOF separation [i.e., $a_{\mathrm{RLOF}}$ in Equation~(\ref{ eq:RLOF })].
The competition between the two effects can result in different evolution
processes. If $a_{\mathrm{RLOF}}$ grows much faster, the NS-WD system will
quickly progress into runaway mass transfer, tidally disrupting the WD on a
dynamical time-scale. Otherwise, the NS-WD system can maintain its stable mass
transfer for a longer time. Many studies have found that the stability of mass
transfer depends heavily on the mass ratio $q$ \citep{1988ApJ...332..193V,
Paschalidis:2009zz, Margalit:2016joe}. They suggest that the mass transfer is
unstable for binaries with $q_{\text {crit }} \gtrsim
0.43$--$0.53$.\footnote{\citet{Bobrick:2017rlo} have recently suggested that
winds from the accreting stream are far more important to the stability, which
brings a much smaller $q_{\text {crit }} \gtrsim 0.20$. Note that such a lower
$q_{\text {crit }}$ value is also in line with the results in
\citet{Kaltenborn:2022vxf}.} Combined with the distribution of $M_{\mathrm{WD}}$
in our consideration (see Figure~\ref{ fig:fbm }), it is safe to say that most
NS-WD systems would finally have a runaway mass-transfer phase, rather than the
stable mass transfer. Given that there is no well-modeled GW waveform for the
NS-WD system during the runaway mass-transfer phase, we define the time when the
WD starts to experience the RLOF as the merger time. We regard this as a
reasonable treatment, especially considering the short tidal-disruption
timescale, $\lesssim {\cal O}(\mathrm{min})$ \citep[see,
e.g.,][]{Kaltenborn:2022vxf} and possibly observable explosive outcomes  for
NS-WD mergers \citep{King:2006nw, Chattopadhyay:2007rx, Metzger:2011zk,
Margalit:2016joe, Margalit:2016hlx, Yang:2022qmy, Zhong:2023zwh}. Note that such
definitions would be very different from those in BNS and NS-BH mergers in many
studies \citep{Kyutoku:2011vz, Shibata:2011jka, Zhu:2020ffa, Liu:2021dcr,
Liu:2022mcd}. Essentially, for the NS-WD mergers in this work, the GW
early-warning detections are actually to offer alerts on the RLOF time. For this
reason, we will only consider the GW signal detectability in the inspiral phase,
when the NS and WD can be regarded as well-separated bodies that gradually
spiral towards one another. More descriptions of GW early-warning detections are
presented in Section~\ref{ sec:warnings }.

\section{GW detection strategy} 
\label{ sec:warnings }

As mentioned in the Introduction, due to the higher sensitive frequency range of
the ground-based GW detectors, they cannot offer alerts as early as the GW
detectors in the decihertz band. Moreover, for space-borne decihertz GW
detectors, NS-WD signals can exist from the start of the mission to their
mergers or even to the end of the mission. Therefore, such GW signals will exist
in decihertz detectors long enough to guarantee a relatively stable parameter
estimation precision, especially for the localization. This means
that the decihertz GW detectors can provide early-warning alerts to other GW and
EM detectors for follow-ups. Following \citet{Liu:2022mcd}, our realistic
detection strategy is performed as follows. Note that we use geometric units
where $G=c=1$ in this section.  

Throughout this paper, we mainly compare the performance on GW early warnings
between two space-borne decihertz GW detectors, DO-OPT and DEC. DO was envisaged
in the ESA's Voyage 2050 call \citep{Sedda:2019uro, Sedda:2021yhn}, where DO-OPT
is the one with more ambitious LISA-like designs.  DEC is a future Japanese
space-borne decihertz GW mission with four independent LISA-like detectors
\citep{{Kawamura:2011zz, Kawamura:2020pcg}}. Parameters of the two detectors are
available in, e.g., \citet{Liu:2021dcr}. In the community, there are more
proposed decihertz detectors, including \add{atom-interferometer-based detectors \citep{Zhao:2021bjw, Baum:2023rwc} and moon-based ones \citep{Jani:2020gnz, LGWA:2020mma, Li2023, Shao2023}. In comparison, the sensitivity curves of DO-OPT and DEC are expected to perform much better than these mentioned detectors in the decihertz band.}

\begin{figure}
    \centering
    \includegraphics[width=8.5cm]{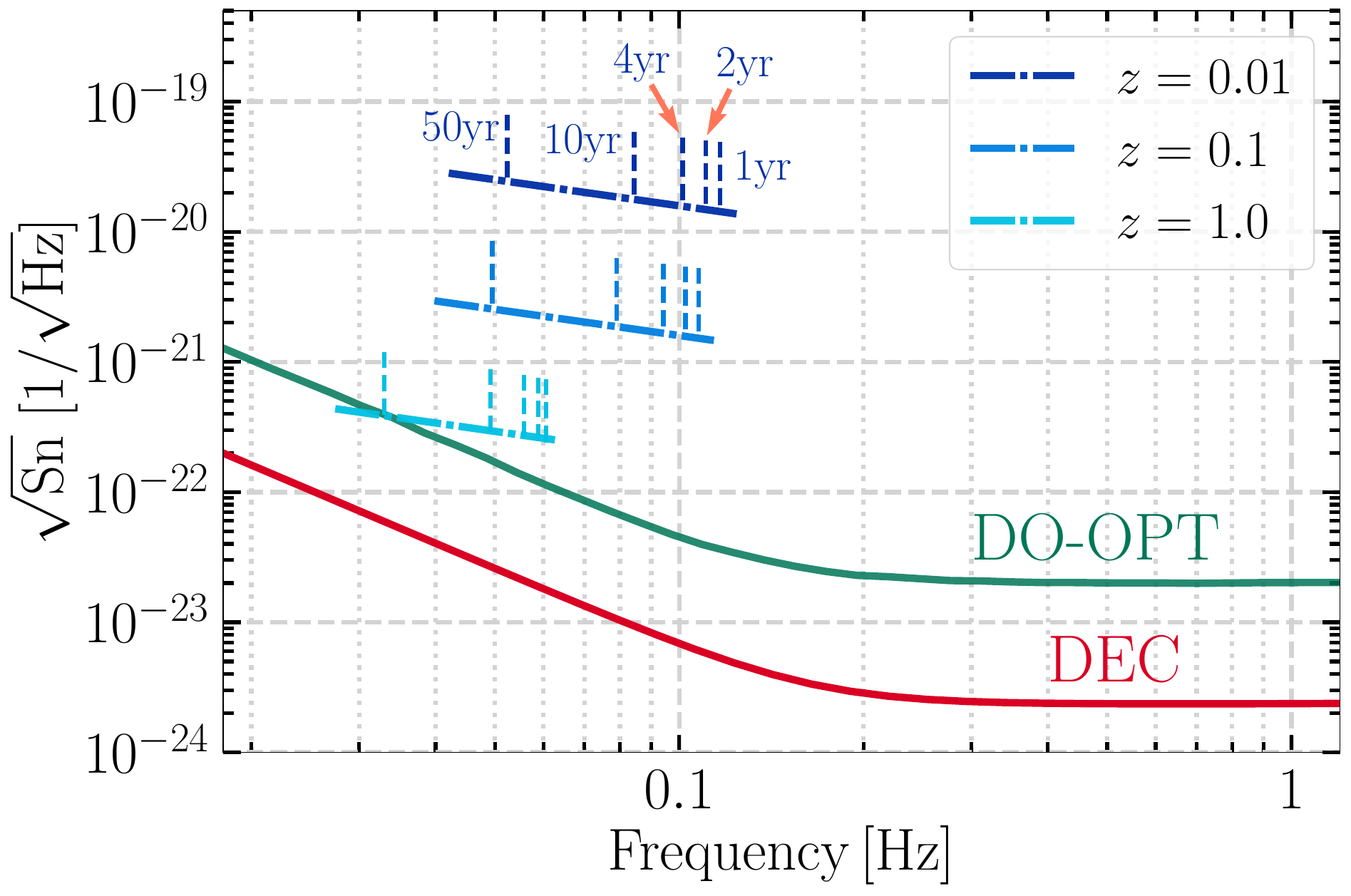}
    \caption{Sensitivity curves of DO-OPT (green) and DEC (red). The simulated
    NS-WD inspiral signals at different redshifts are plotted with blue
    dashed-dotted lines. Taking $M_\mathrm{NS} = M_\mathrm{WD} =
    1.1\,\mathrm{M}_{\odot}$ as an example, source signals are plotted with a
    duration of $\sim 100\,\mathrm{yr}$. For each source, the short dashed
    vertical lines mark the time before RLOF occurs (i.e., the rightmost
    endpoint). More descriptions of the sky-averaged effective noise
    $\sqrt{S_\mathrm{n}}$ for various detectors can be found in
    \citet{Liu:2020nwz}.} 
    \label{ fig:Detector }
\end{figure}

We define $t_{\mathrm{c}_{0}}$ as the NS-WD's time to merge since the start of
the observation with decihertz GW detectors. To obtain the yearly detection
rates, we only use early-warning NS-WD samples that will merge in
1--$2\,\mathrm{yr}$ since the mission begins (e.g., $1\,\mathrm{yr} \leq
t_{\mathrm{c}_{0}} \leq 2\,\mathrm{yr}$). The sources that merge within 1\,yr
($t_{\mathrm{c}_{0}} < 1\,\mathrm{yr}$) are discarded. This is because their GW
signals only stay shortly in the decihertz detectors, and not enough information
is accumulated to obtain precise parameter estimations. \citet{Liu:2022mcd} have
shown that decihertz detectors have poor performances for BNSs that merge within
1\,yr, which also applies to our NS-WD mergers. Considering that NS-WD mergers
typically produce signals at around 0.1\,Hz, there will only be a slight
frequency chirp effect in the decihertz band (see Figure~\ref{ fig:Detector }).
This means that the choice of the maximum value of $t_{\mathrm{c}_{0}}$ will
have minor effects on the yearly detection results as long as it is not too
long.\footnote{Note that people usually set a 4-yr mission time for DO-OPT and
DEC. For sources that only inspiral within the whole 4-yr observational span
(e.g., $t_{\mathrm{c}_{0}}>4\,\mathrm{yr}$), \citet{Liu:2022mcd} have shown that
their timing accuracies have a sharp decreasing trend, which is unfavorable for
GW early-warning detections. In contrast, sources with $1\,\mathrm{yr} \leq
t_{\mathrm{c}_{0}} \leq 4\,\mathrm{yr}$ are in general distributed uniformly in
time \citep[see Table~1 in][]{Liu:2022mcd}.} The above discussions explain
why we only consider  NS-WD mergers with $1\,\mathrm{yr} \leq t_{\mathrm{c}_{0}}
\leq 2\,\mathrm{yr}$.

Following \citet{Liu:2022mcd}, we generate NS-WD mergers weekly according to the
population models. Then we calculate their SNRs. If SNR > 8, we claim the
detection and calculate parameter precisions using the Fisher matrix
\citep{Finn:1992wt, Cutler:1994ys}. As a fast way to estimate parameter
statistical errors, a Fisher information matrix (FIM) shows the Cramer–Rao bound
of parameters. In other words, a FIM tells us how precisely we can determine the
model parameters for events with a high SNR and Gaussian noises (see
\citet{Wang:2022kia} for extensions of the FIM). \add{Note that the choice of SNR threshold may vary depending on the specific goals of a study, which deserves more detailed analyses.} We follow the settings in
\citet{Liu:2021dcr} with slight revisions. In total nine system parameters are
used in the FIM, collectively denoted as 
\begin{equation}
	\boldsymbol{\Xi}=\left\{\mathcal{M}_{z}, \eta, t_{\mathrm{c}},
	\phi_{\mathrm{c}}, D_{\mathrm{L}}, \bar{\theta}_{\mathrm{S}},
	\bar{\phi}_{\mathrm{S}}, \bar{\theta}_{\mathrm{L}},
	\bar{\phi}_{\mathrm{L}}\right\} \,.
\end{equation}
In $\boldsymbol{\Xi}$, $\eta \equiv M_{\mathrm{NS}} M_{\mathrm{WD}}
/\left(M_{\mathrm{NS}}+M_{\mathrm{WD}}\right)^{2}$ is the symmetric mass ratio
for each NS-WD merger system; $\mathcal{M}_{z} \equiv (1+z)
\left(M_{\mathrm{NS}}+M_{\mathrm{WD}}\right) \eta^{3 / 5}$ is the
detector-frame chirp mass; $t_{\mathrm{c}}$ and $\phi_{\mathrm{c}}$ are the time
and orbital phase at coalescence, respectively; and
$\left\{\bar{\theta}_{\mathrm{S}}, \bar{\phi}_{\mathrm{S}},
\bar{\theta}_{\mathrm{L}}, \bar{\phi}_{\mathrm{L}}\right\}$ are the source
direction and angular momentum direction in the Solar system barycentric frame
\citep[see Figure~1 in][]{Liu:2020nwz}. Note that in generating sources, we
adopt $\cos \bar{\theta}_{\mathrm{S}} \in \mathcal{U}(-1,1)$, $\cos
\bar{\theta}_{\mathrm{L}} \in \mathcal{U}(-1,1)$, $\bar{\phi}_{\mathrm{S}} \in
\mathcal{U}(0,2 \pi)$, and $\bar{\phi}_{\mathrm{L}} \in \mathcal{U}(0,2 \pi)$,
where $\mathcal{U}(\cdot\,,\cdot)$ denotes a uniform distribution.

After performing the FIM calculation, we obtain the angular resolution $\Delta
\Omega$ via \citep{Cutler:1997ta, Barack:2003fp},
\begin{equation}
    \Delta \Omega=2 \pi \sqrt{\left(\Delta \bar{\mu}_{\mathrm{S}} \Delta
    \bar{\phi}_{\mathrm{S}}\right)^{2}-\left\langle\delta \bar{\mu}_{\mathrm{S}}
    \delta \bar{\phi}_{\mathrm{S}}\right\rangle^{2}} \,,
    \label{ eq:delta omega }
\end{equation}
where $\Delta \bar{\mu}_{\mathrm{S}}$ and $\Delta \bar{\phi}_{\mathrm{S}}$ are
the root-mean-square errors of $\bar{\mu}_{\mathrm{S}}$ and
$\bar{\phi}_{\mathrm{S}}$ with $\bar{\mu}_{\mathrm{S}} \equiv \cos
\bar{\theta}_{\mathrm{S}}$, and $\left\langle\delta \bar{\mu}_{\mathrm{S}}
\delta \bar{\phi}_{\mathrm{S}}\right\rangle$ is the covariance of
$\bar{\mu}_{\mathrm{S}}$ and $\bar{\phi}_{\mathrm{S}}$. In this work, our
attention focuses on the estimations of $\Delta \Omega$ and $\Delta
D_{\mathrm{L}}$.

For the FIM in the frequency domain, we follow \citet{Kang:2022nmz}  with slight
revisions. We set the integration limit to be $f_{\text {in}}$ and $f_{\text
{out}}$. In an idealized setting of two point particles, assuming a circular
Keplerian orbit with quadrupolar GW damping, the GW frequency will formally
diverge at a finite value of time, $t_{\text {div}}$. However, for a real NS-WD
system, we have noted previously in Section~\ref{ sec:mdistri } that the merger
time $t_{\mathrm{c}_{0}}$ corresponds to the time when the WD starts to
experience the RLOF. Thus the upper cutoff GW frequency in this work is actually
at the critical RLOF separation, which can be calculated, to a sufficient
precision, by Kepler’s third law,
\begin{equation}
f_{\max } \simeq \frac{1}{\pi (1+z)}
\sqrt{\frac{M_{\mathrm{NS}}+M_{\mathrm{WD}}}{a_{\mathrm{RLOF}}^{3}}} \,.
\label{ eq:f_max }
\end{equation}
Note that the $(1+z)$ foctor is included in the denominator due to the
cosmological time dilation. With a known $f_{\max }$ value for each NS-WD
system, we can then calculate the remaining time to divergence
\citep{Maggiore:2007ulw},
\begin{equation}
\begin{aligned} 
    t_{\mathrm{RLOF}} =t_{\text {div}} - t_{\mathrm{c}_{0}} = 5
    \mathcal{M}_{z}^{-5 / 3} (8 \pi f_{\max })^{-8/3 } \,.
\end{aligned} 
\label{ eq:t_RLOF }
\end{equation}
For a NS-WD system to merge in time $t_{\mathrm{c}_{0}}$, we thus have
\begin{align}
	f_{\text{in}} &=\left[(t_{\mathrm{c}_{0}}+t_{\mathrm{RLOF}}) / 5\right]^{-3
	/ 8} \mathcal{M}_{z}^{-5 / 8} / 8 \pi \,, \\
	f_{\text {out}} &= ~ {\left[(t_{\mathrm{e}}+t_{\mathrm{RLOF}}) /
	5\right]^{-3 / 8} \mathcal{M}_{z}^{-5 / 8} / 8 \pi} \,,
\end{align}
with $t_{\text {e}}$ the early-warning time before the merger. The early-warning
time $t_{\text {e}}$ is remarkably significant to the GW early-warning
detections of NS-WD mergers, given the short tidal-disruption timescale and
possibly observable EM transients during the runaway mass-transfer phase. We
will present the results with $t_{\mathrm{e}} = 1\,\mathrm{d}$ in detail. More
descriptions of the GW waveform construction and parameter estimation method
were discussed in \citet{Liu:2021dcr}.

\begin{table*}
    \renewcommand\arraystretch{2}
    \centering
    \caption{Yearly detection numbers and percentages (in brackets) of NS-WD
    mergers for different population models with DO-OPT and DEC. We assume an
    early-warning time $t_\mathrm{e} = 1\,\mathrm{d}$. The SNR threshold is set
    to be 8. Note that we only list results with $\dot{\rho}_{0} =
    390\,\mathrm{Gpc}^{-3}\,\mathrm{yr}^{-1}$. Readers can rescale in the way as
    for the total simulated numbers in Table~\ref{ tab:Total numbers }.}
    \setlength{\tabcolsep}{1.1cm}{\begin{tabular}{c c c c c}
    \toprule
    \toprule
    \vspace{-3.5em}\\
    GW Detector & \multicolumn{4}{c}{Population Model}\\  
    \cline{2-5}
    & A    & B    & C    & D\\
    \cline{1-5}
    \multirow{2}{30pt}{\shortstack{DO-OPT}} 
    & 1915    
    & 1637    
    & 1722    
    & 1489\vspace{-1em}\\  
    & (0.34\,\%)            
    & (0.40\,\%)          
    & (0.35\,\%)            
    & (0.40\,\%)\\
    \cline{1-5}
    \multirow{2}{18pt}{\shortstack{DEC}}
    & 46049    
    & 36134    
    & 41272    
    & 33314\vspace{-1em}\\   
    & (8.25\,\%)           
    & (8.75\,\%)  
    & (8.34\,\%)         
    & (8.86\,\%)\vspace{0em}\\
    \bottomrule
    \end{tabular}}
    \label{ tab:detection rates }
\end{table*}

Finally, a word on the tidal deformability included in our GW waveform. At the
late stages of the inspiral, the quadrupolar tidal field $\mathcal{E}_{i j}$ of
one compact component would induce a quadrupole moment $Q_{i j}$ to the other.
To the leading order in the adiabatic approximation, $\mathcal{E}_{i j}$ and
$Q_{i j}$ are related by a linear response function, $Q_{i j}=-\lambda
\mathcal{E}_{i j}$, where $\lambda$ is the tidal Love number
\citep{Hinderer:2007mb}. The static tidal effect enters the GW phase at the 5-th
post-Newtonian order\footnote{Post-Newtonian order represents the power of
velocity squared relative to the leading Newtonian order for point-particle
binary motion.} through the dimensionless binary tidal deformability
$\tilde{\Lambda}$ \citep{Flanagan:2007ix, Hinderer:2007mb, Favata:2013rwa},
\begin{equation}
\tilde{\Lambda}=\frac{16}{13} \frac{\left(M_\mathrm{NS}+12 M_\mathrm{WD}\right)
M_\mathrm{NS}^{4} \Lambda_\mathrm{NS}+\left(M_\mathrm{WD}+12
M_\mathrm{NS}\right) M_\mathrm{WD}^{4}
\Lambda_\mathrm{WD}}{\left(M_\mathrm{NS}+M_\mathrm{WD}\right)^{5}} \,,
\label{ eq:Lambda_tilde }
\end{equation}
where $\Lambda_\mathrm{NS}$ and $\Lambda_\mathrm{WD}$ correspond respectively to
the dimensionless tidal deformability of the NS and the WD in a NS-WD binary
system. With $\Lambda_\mathrm{WD} \gg \Lambda_\mathrm{NS}$\footnote{For cold,
slowly rotating WDs, we can see in Figure~2 of \citet{Wolz:2020sqh} that
$\Lambda_\mathrm{WD}$ is of the order of $10^{13}$--$10^{21}$. However, in
comparison, $\Lambda_\mathrm{NS}$ is $\lesssim \mathcal{O}(10^{3})$ in GW170817
\citep{LIGOScientific:2017vwq}.} we can simplify Equation~(\ref{ eq:Lambda_tilde })
to
\begin{equation}
\tilde{\Lambda} \simeq \frac{\left(M_\mathrm{WD}+12 M_\mathrm{NS}\right)
M_\mathrm{WD}^{4}
\Lambda_\mathrm{WD}}{\left(M_\mathrm{NS}+M_\mathrm{WD}\right)^{5}} \,.
\label{ eq:Lambda_tilde_new }
\end{equation}

We follow \citet{Wolz:2020sqh} to obtain $\Lambda_\mathrm{WD}$ via the universal relation,
\begin{equation}
\ln \Lambda_\mathrm{WD}=2.02942+2.48377 \ln \bar{I}_\mathrm{WD} \,,
\label{ eq:Lambda }
\end{equation}
where $\bar{I}_\mathrm{WD}=I_\mathrm{WD} / M_\mathrm{WD}^{3}$ is the
dimensionless moment of inertia of the WD, which can be calculated with another
fitting universal relation,
\begin{equation}
\begin{aligned} 
    \ln \bar{I}_\mathrm{WD} = & \ 24.7995-39.0476
    \left(\frac{M_{\mathrm{WD}}}{1\,\mathrm{M}_{\odot}}\right) + 95.9545
    \left(\frac{M_{\mathrm{WD}}}{1\,\mathrm{M}_{\odot}}\right)^{2} \\ & -138.625
    \left(\frac{M_{\mathrm{WD}}}{1\,\mathrm{M}_{\odot}}\right)^{3}+98.8597
    \left(\frac{M_{\mathrm{WD}}}{1\,\mathrm{M}_{\odot}}\right)^{4}-27.4
    \left(\frac{M_{\mathrm{WD}}}{1\,\mathrm{M}_{\odot}}\right)^{5}\,.
\end{aligned}
\label{ eq:bar_I }
\end{equation}

Note that although many theoretical studies have discussed the effects of mass
transfer and dynamical tides for WDs \citep{Lai:2011xp, Kremer:2017xrg,
Tauris:2018kzq, McNeill:2019rct, Kuns:2019upi, Lau:2022ejf}, there are still
many uncertainties and unsettled problems, especially on the GW waveform
construction with mass transfer. In addition, many studies have focused on the
low-mass WD donors (i.e., He WDs) or BWD systems, which can be different from
NS-WD binaries in our consideration. In the following sections, we mainly focus
on the estimation of the accuracy of distance $\Delta D_{\mathrm{L}}$ and
angular resolution $\Delta \Omega$. These two quantities can be extracted from
the inspiral chirp signal alone. We leave other complex effects involving  WDs
for future studies.

\section{Result} 
\label{ sec:Results }

\begin{figure*}
    \centering
    \includegraphics[width=8.2cm]{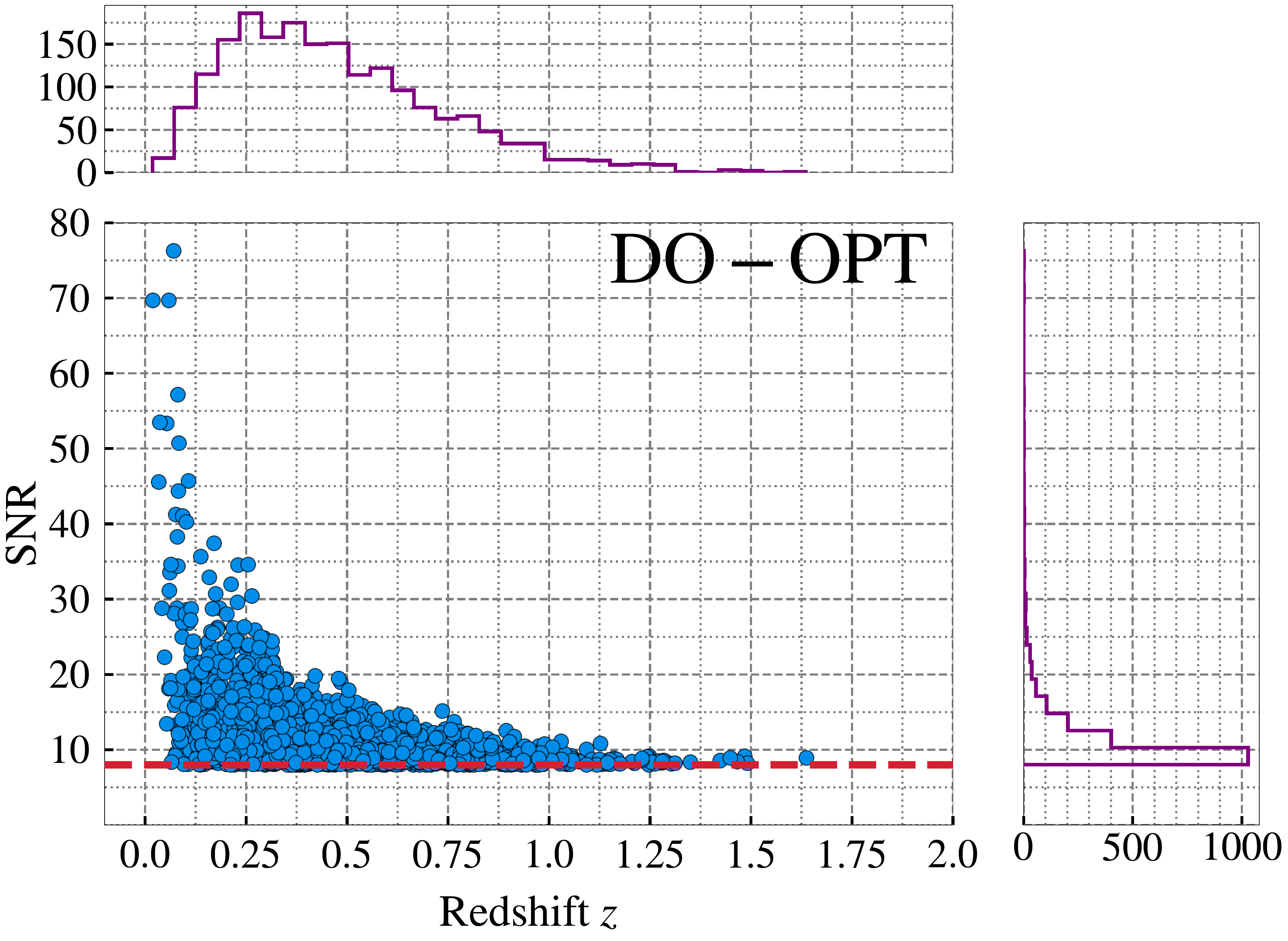}
    \hspace{2.5em}
    \includegraphics[width=8.2cm]{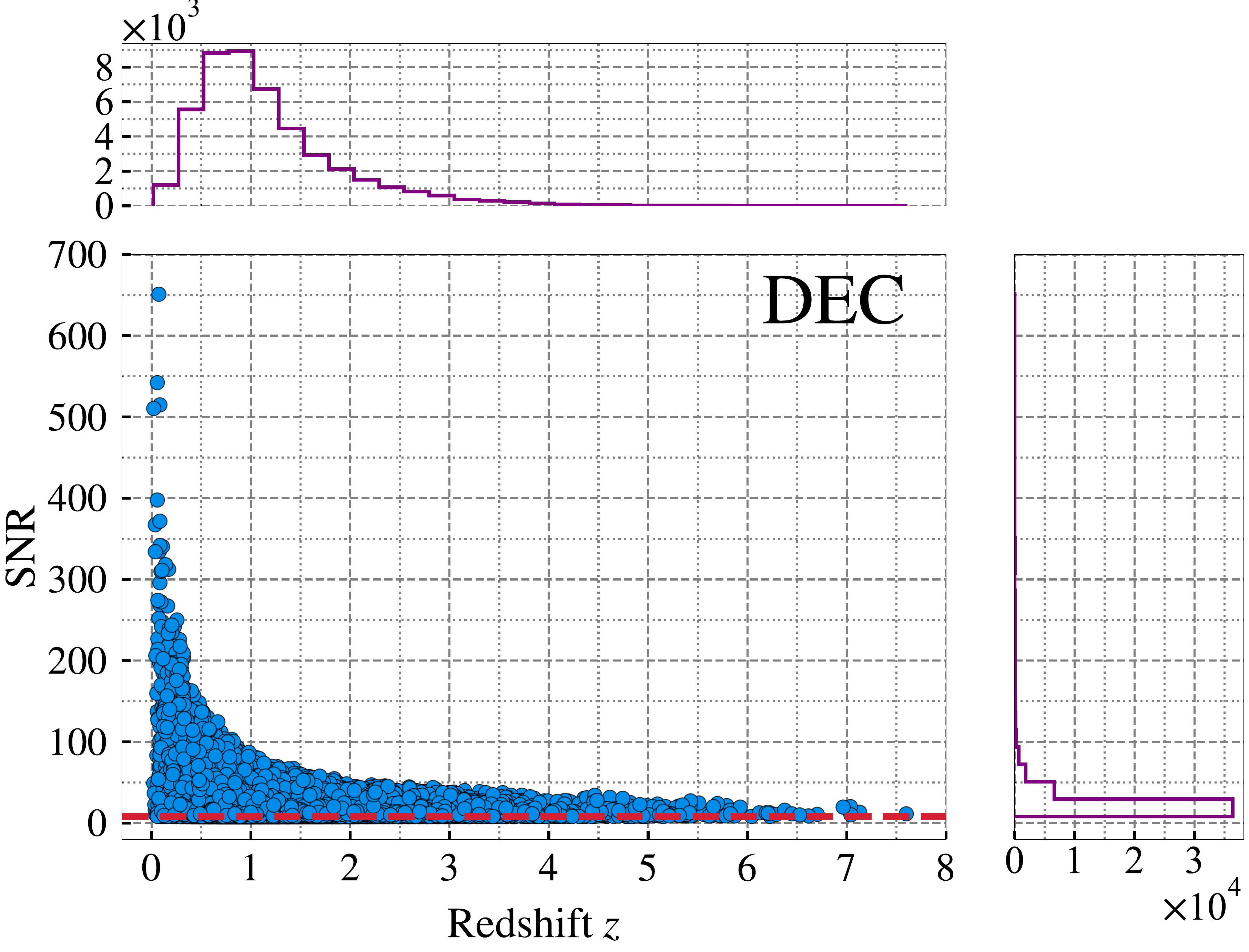}
    \caption{SNR as a function of redshift for yearly detectable NS-WD mergers
    for DO-OPT (left) and DEC (right). We only show the population Model A as an
    example. Dashed horizontal red lines correspond to the detection threshold
    $\mathrm{SNR} = 8$. In each plot, the purple histograms show the redshift
    and SNR distributions in upper and right panels, respectively.} 
    \label{ fig:SNR_z }
\end{figure*}

In Table~\ref{ tab:detection rates } we first summarize yearly detection numbers
of NS-WD mergers for the four population models with DO-OPT and DEC. We also
list corresponding percentages in brackets, showing the ratio to the total
number of yearly simulated NS-WD mergers (cf Table~\ref{ tab:Total numbers }).
The SNR threshold is set to 8. Combining all the results shown in Table~\ref{
tab:detection rates }, we conclude that DEC performs much better than DO-OPT on
GW early-warning detections as a whole, with the yearly detection number in the
range of $\sim (3.3\mbox{--}4.6) \times 10^{4}$. The detection number for DO-OPT
is in the range of $\sim (1.5\mbox{--}1.9) \times 10^{3}$, which is
approximately 20 times fewer than DEC. Given that DEC has a lower noise level
(see Figure~\ref{ fig:Detector }), it is reasonable for DEC to show better
detection abilities, especially for detection at higher redshift, which is
clearly shown in Figure~\ref{ fig:SNR_z }. Taking the population Model A as an
example, Figure~\ref{ fig:SNR_z } shows that DO-OPT can detect early-warning
sources up to $z \lesssim 1.7$, while DEC could reach $z \lesssim 7.6$.
Moreover, the detectable NS-WD mergers for DEC are clustered at $z \simeq 0.8$,
consistent with the peak of the redshift factor $f(z)$ for the population Model
A (see Figure~\ref{ fig:fz }). In comparison, due to its limited detection
abilities, most detectable NS-WD mergers for DO-OPT are clustered around $z
\simeq 0.25$, which is much closer than those of DEC. Similarly, for the maximum
SNR value, Figure~\ref{ fig:SNR_z } shows that DEC could reach $\mathrm{SNR}
\simeq 650$, while DO-OPT only reaches $\mathrm{SNR} \simeq 75$.

\begin{figure*}
    \centering
    \includegraphics[width=8.2cm]{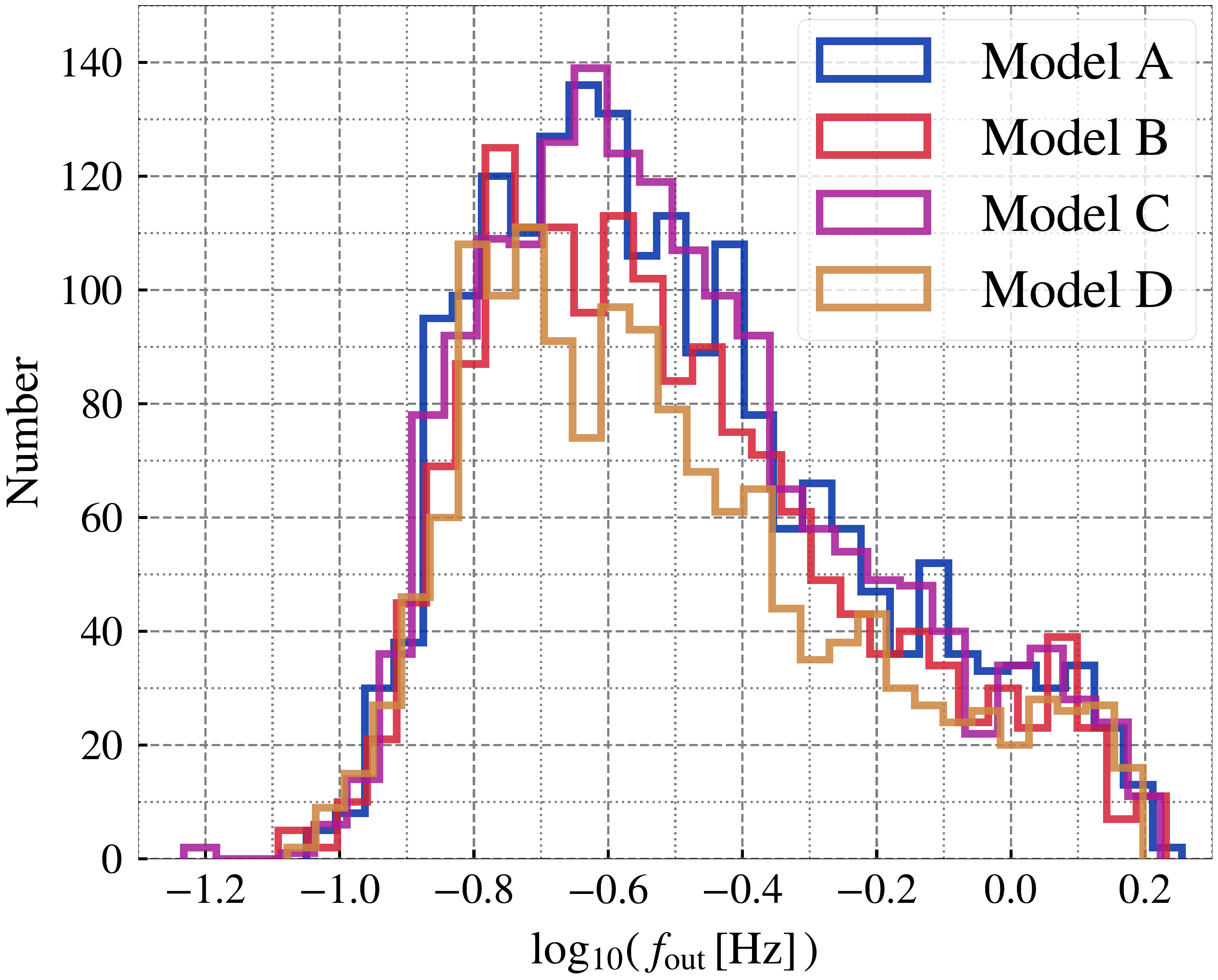}
    \hspace{2.5em}
    \includegraphics[width=8.2cm]{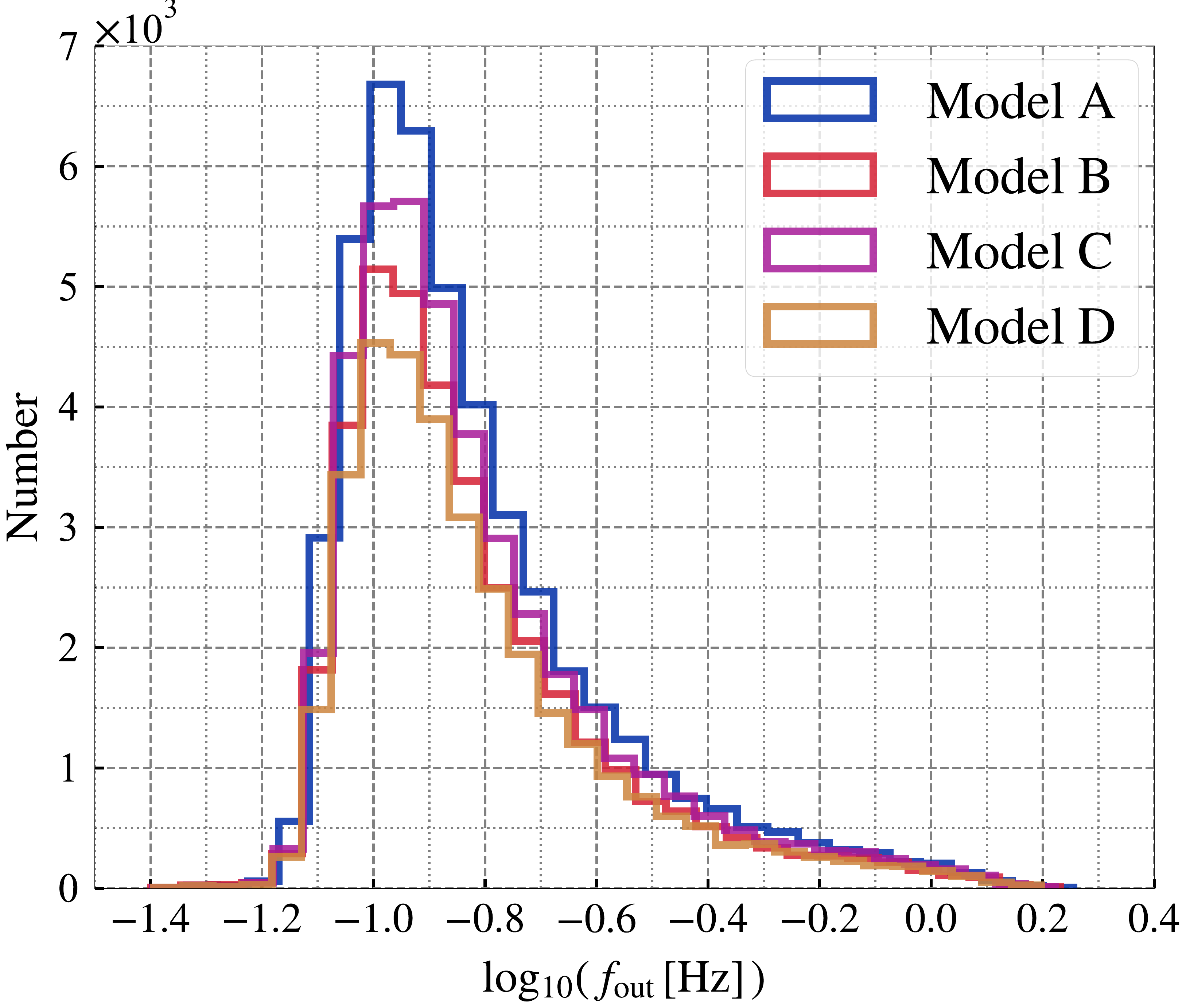}
    \caption{Distributions of $f_{\text {out}}$ (defined in Section~\ref{
    sec:warnings }) for detectable NS-WD mergers in DO-OPT (left) and DEC
    (right). The total numbers of detectable merger events are listed in
    Table~\ref{ tab:detection rates }. We set the early-warning time to be
    $t_\mathrm{e} = 1\,\mathrm{d}$. Different colours correspond to different
    population models.} 
    \label{ fig:fout_counts }
\end{figure*}

As mentioned in Section~\ref{ sec:warnings }, GW early-warning detections of NS-WD
mergers are essentially to offer alerts on the RLOF time. Assuming an
early-warning time $t_\mathrm{e} = 1\,\mathrm{d}$ \footnote{\add{For the merger time uncertainty $\Delta t_{\mathrm{c}}$, we find that 90\% of the yearly detectable NS-WD mergers (see Table~\ref{ tab:detection rates }) could reach a timing accuracy of $\Delta t_{\mathrm{c}} < \mathcal{O}(10^{4})\,\mathrm{s}$ level, regardless of the population models and decihertz GW detectors. Even considering the extra exposure time and slew time for EM follow-up facilities, we still regard this as a crude but acceptable treatment to adopt an early-warning time $t_\mathrm{e} = 1\,\mathrm{d}$ in this work.}} for each NS-WD merger event,
we calculate the integration upper limit $f_{\text {out}}$ with the known
$f_{\max }$ at the critical RLOF separation. We plot in Figure~\ref{
fig:fout_counts } the distributions of $f_{\text {out}}$ for detectable NS-WD
mergers observed by DO-OPT and DEC. Different colours correspond to different
population models. As clearly shown in Figure~\ref{ fig:fout_counts }, the
majority of total detectable NS-WD mergers for DEC would cluster at $f_{\text
{out}} \simeq 0.1\,\mathrm{Hz}$. In comparison, the peak of $f_{\text {out}}$
distribution for DO-OPT would be a little higher in a wider rage
($0.16\mbox{--}0.25\,\mathrm{Hz}$). These could be explained by the cosmological
redshift effect, given that DEC detect more high-$z$ events.

\begin{figure*}
    \centering
    \includegraphics[width=8.2cm]{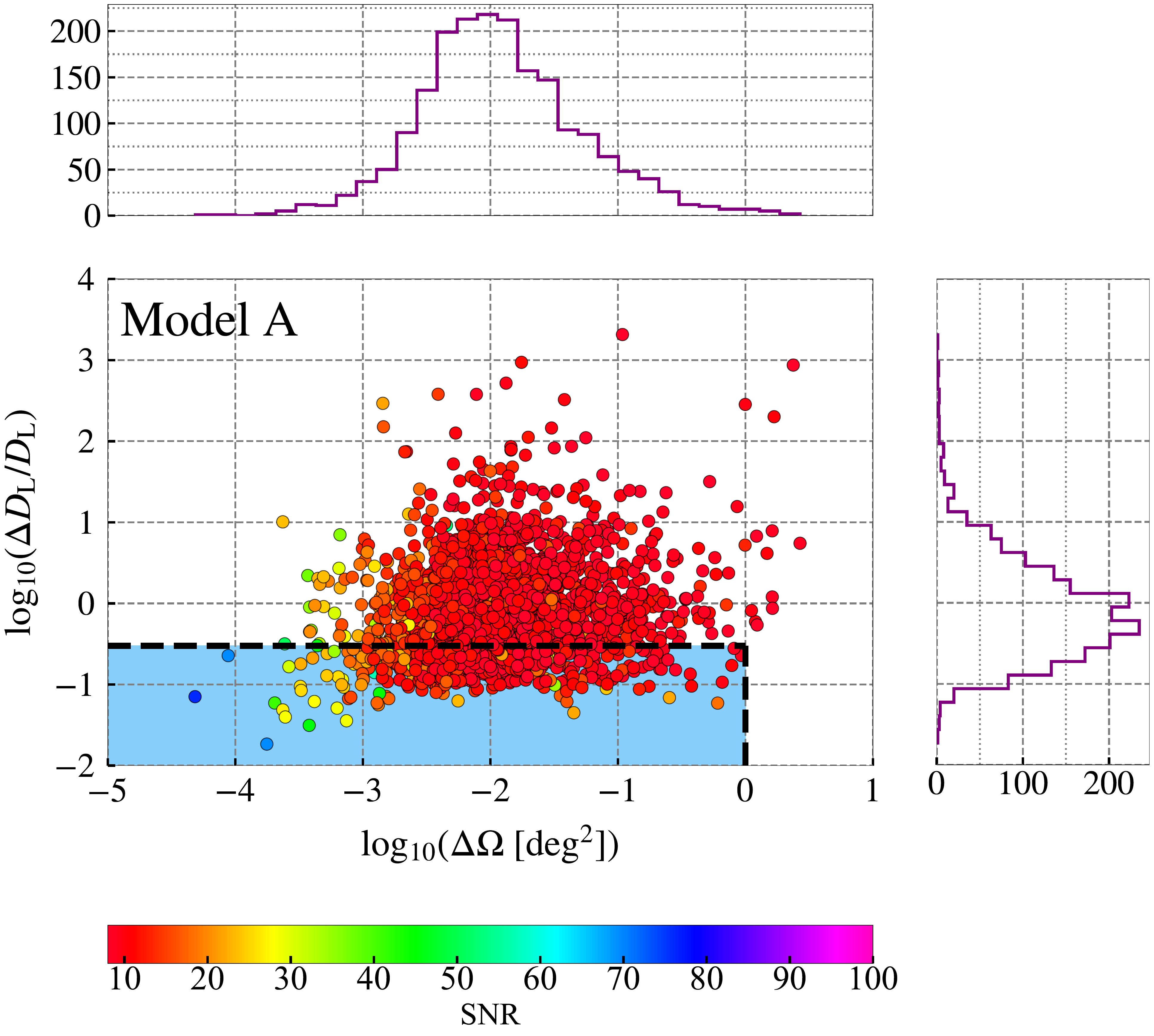}
    \hspace{2.5em}
    \includegraphics[width=8.2cm]{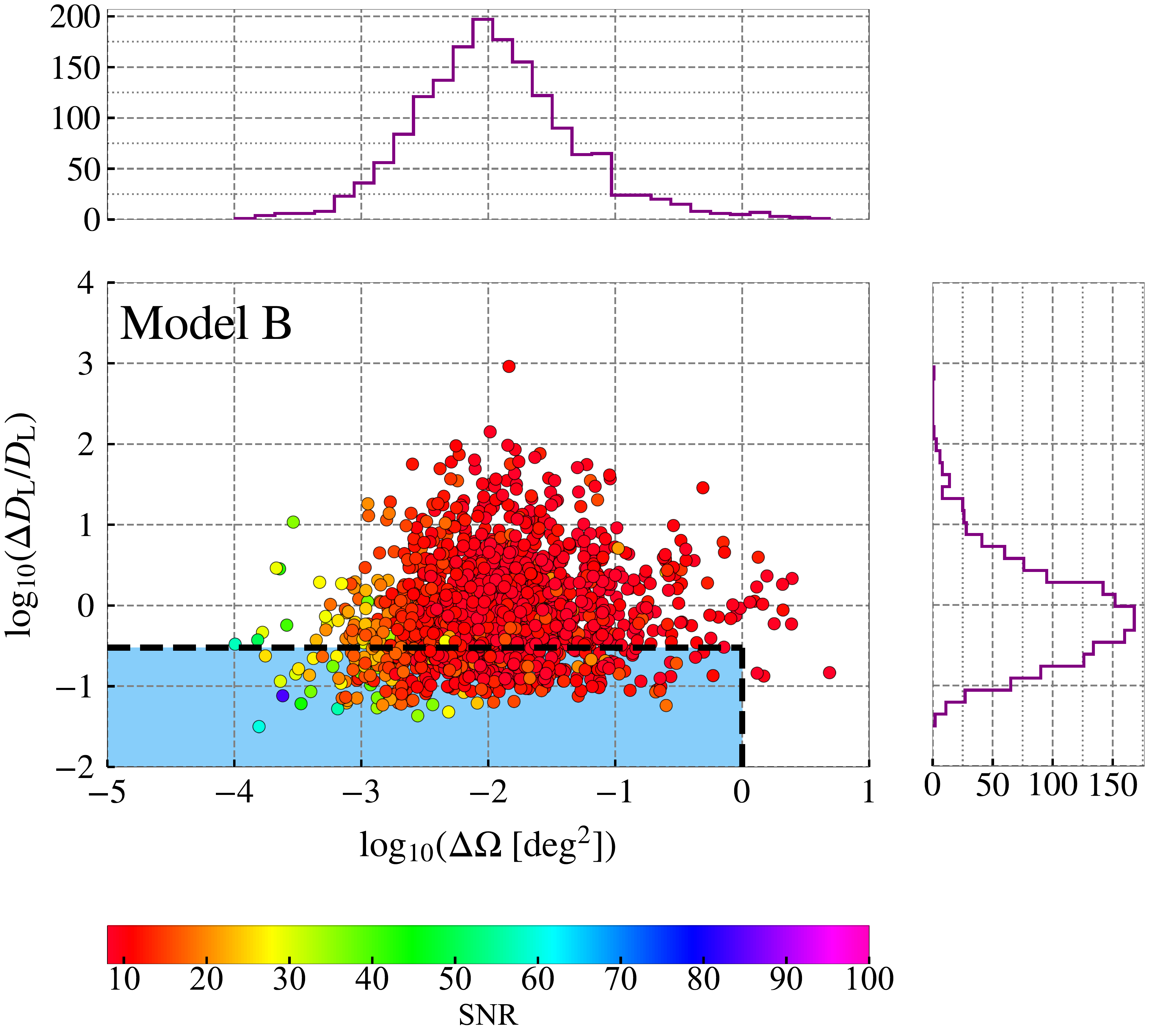}\vspace{1em}\\
    \includegraphics[width=8.2cm]{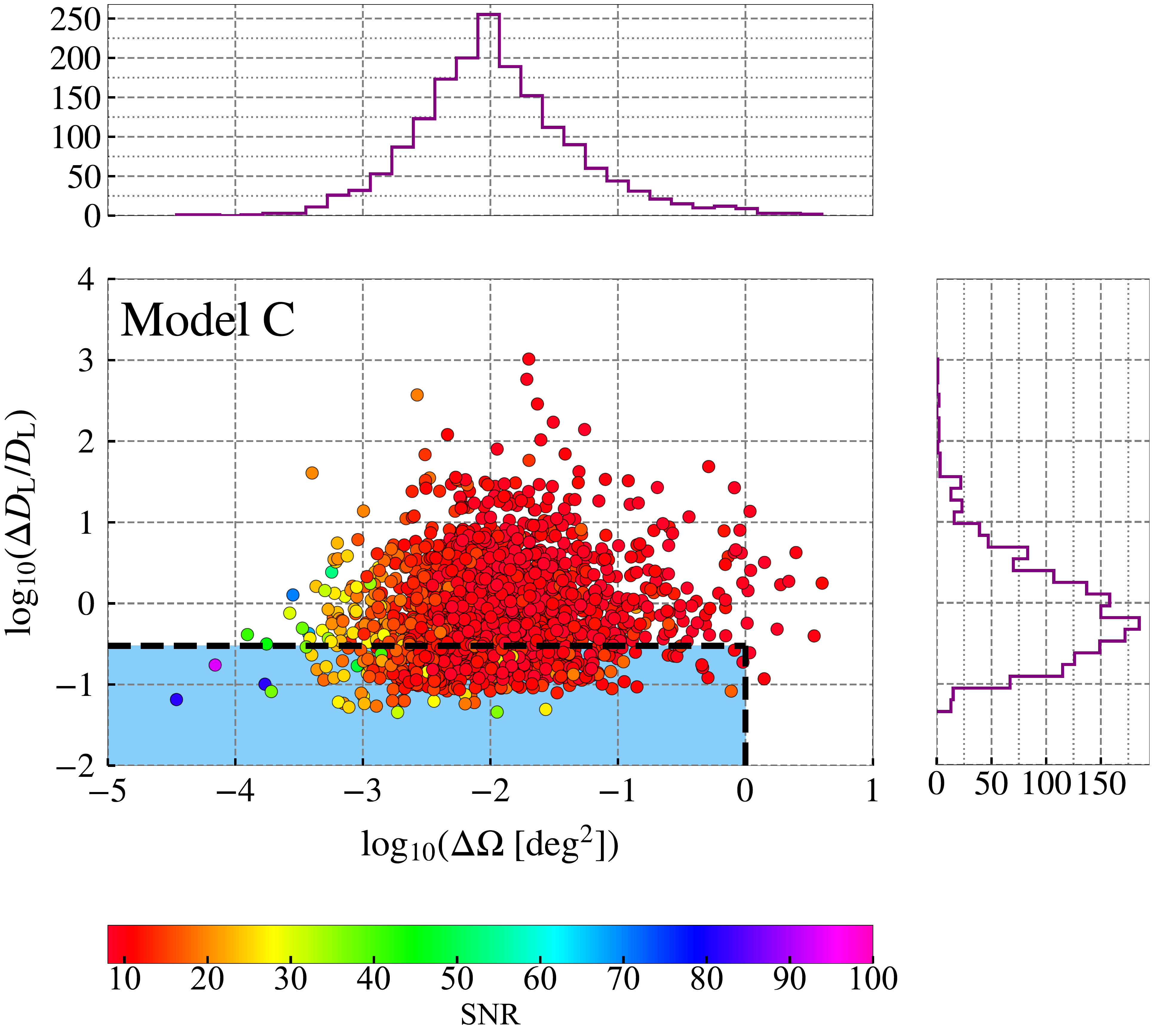}
    \hspace{2.5em}
    \includegraphics[width=8.2cm]{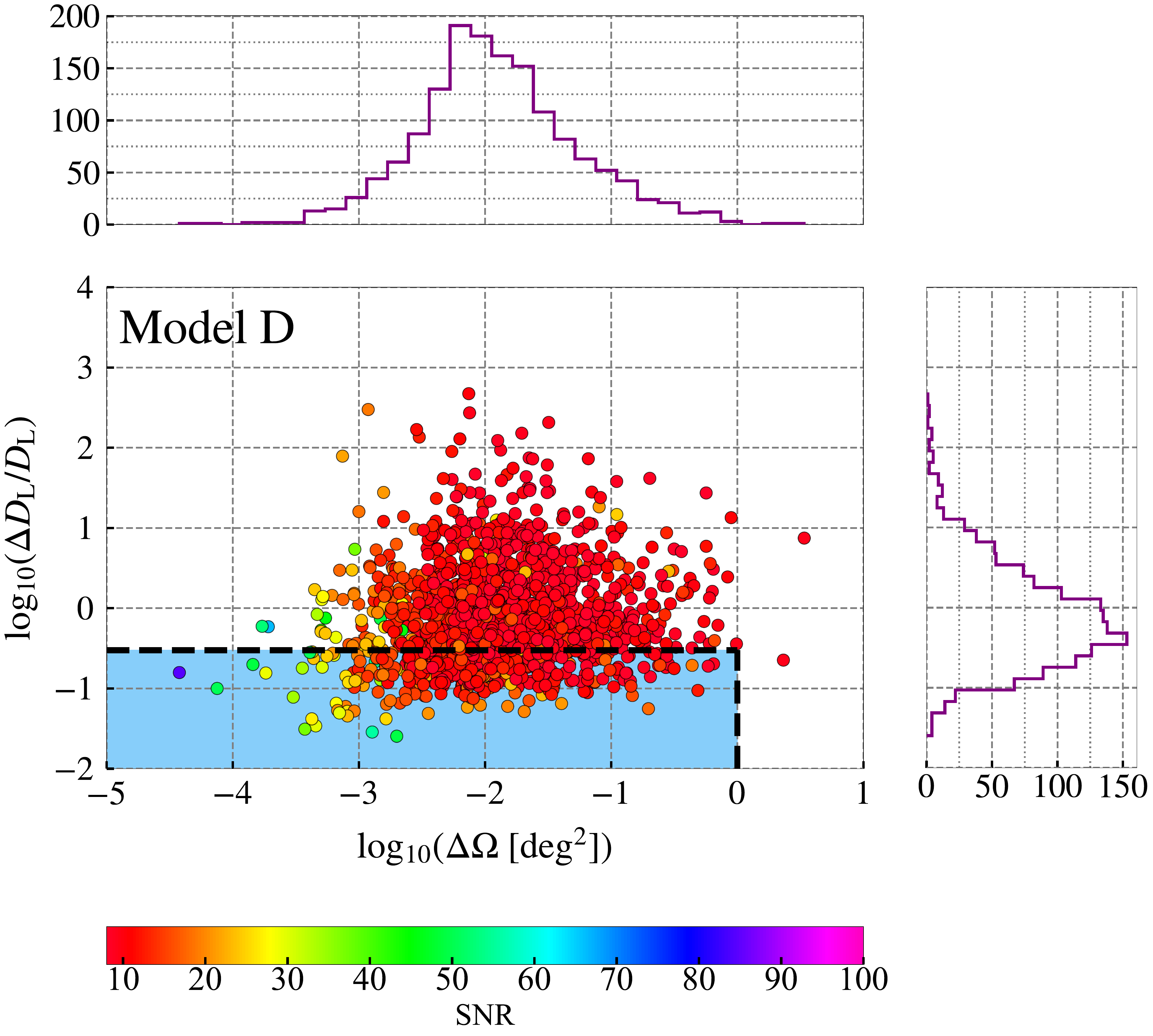}
    \caption{${\Delta D_{\mathrm{L}}}/{D_{\mathrm{L}}}$ versus $\Delta \Omega$
    for yearly GW early-warning samples with DO-OPT for different population
    models. The colour represents the SNR of each merger event. In each plot,
    the purple histograms show the $\Delta \Omega$ and $\Delta
    D_{\mathrm{L}}/D_{\mathrm{L}}$ distributions in the upper and right panels,
    respectively. The black dashed lines denote the dividing lines used to
    select the GS set in the shaded blue region (see Section~\ref{ sec:Results
    }).} 
    \label{ fig:DL_DOmega_DO }
\end{figure*}

\begin{figure*}
    \centering
    \includegraphics[width=8.2cm]{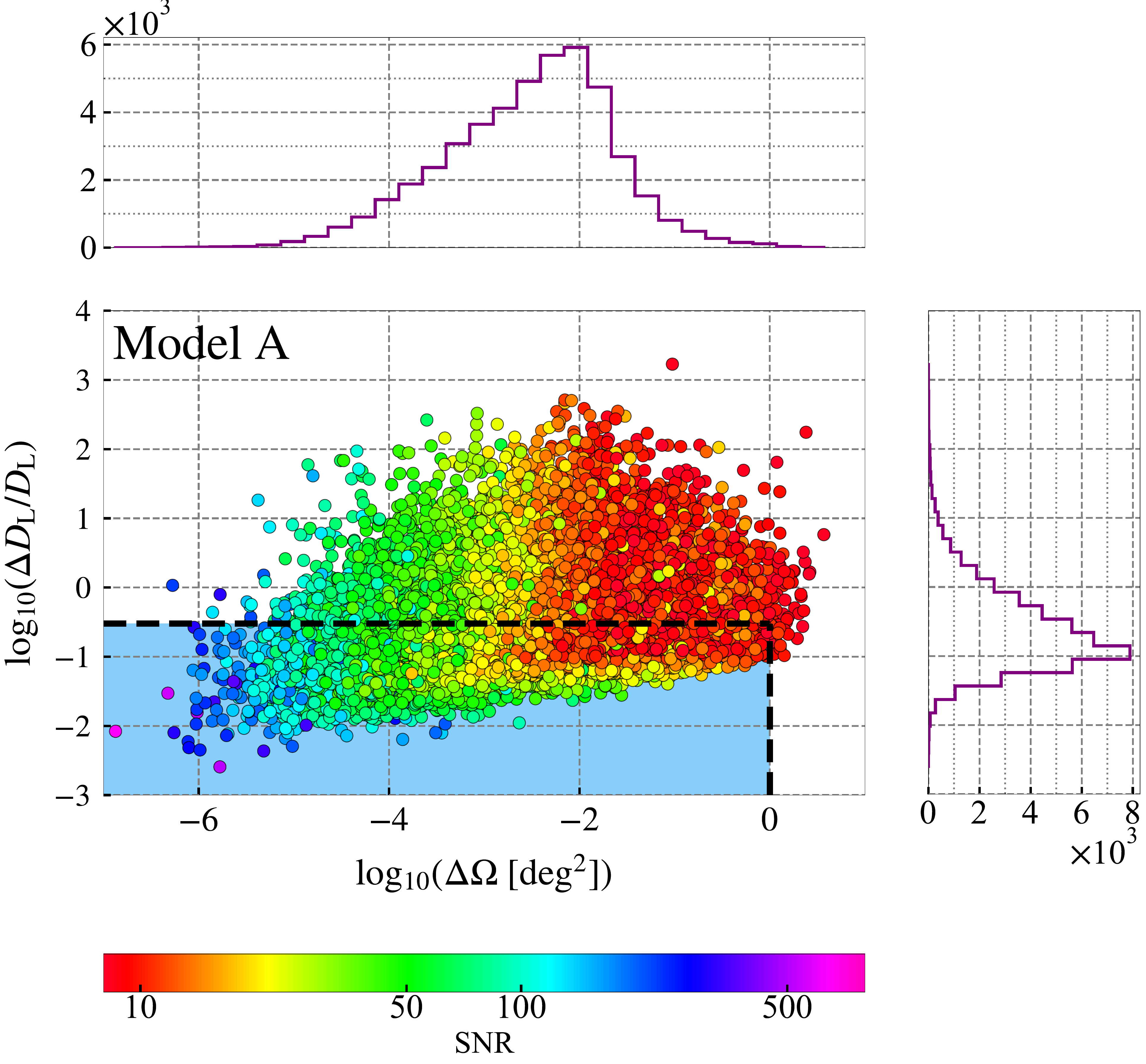}
    \hspace{2.5em}
    \includegraphics[width=8.2cm]{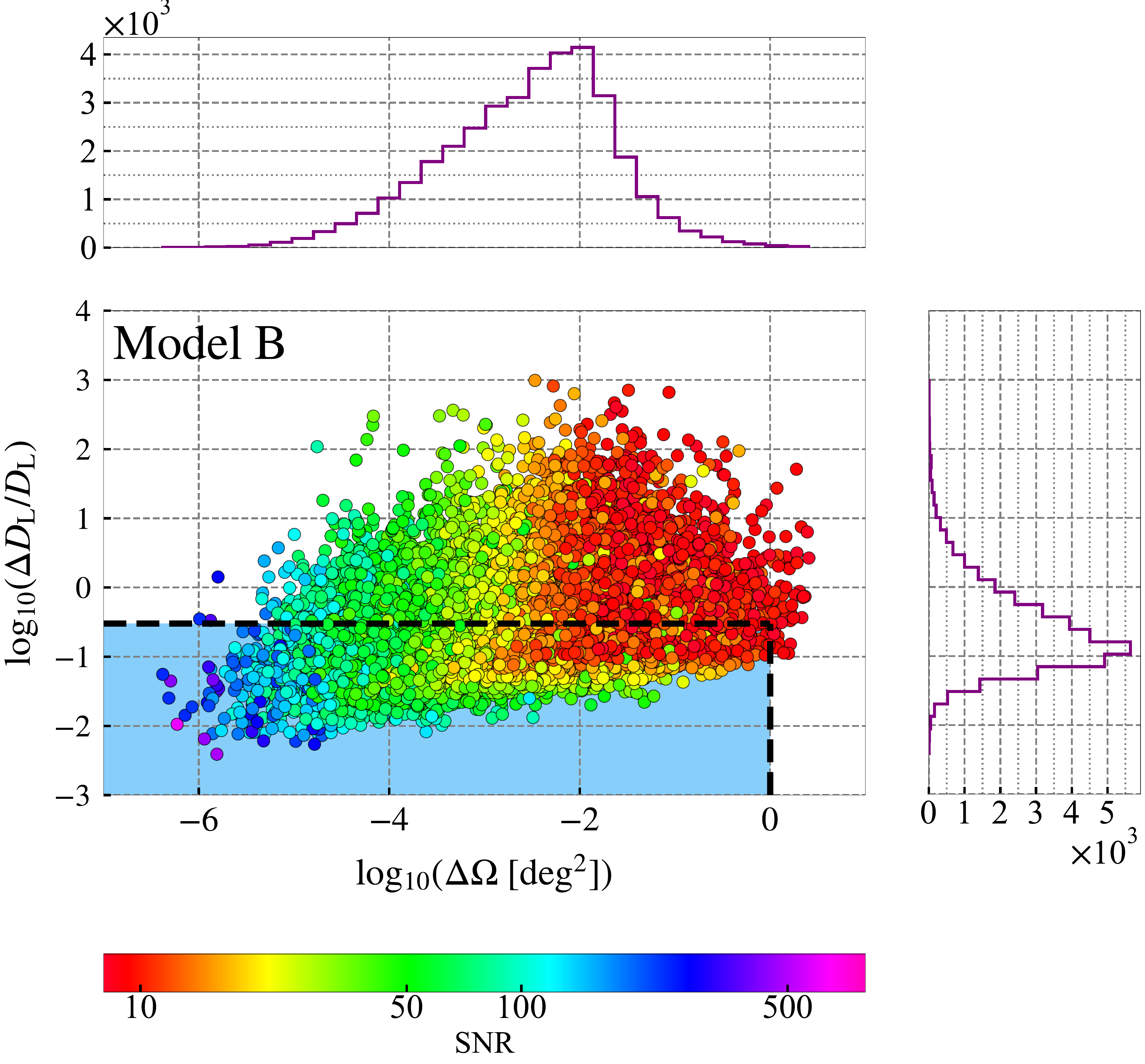}\vspace{1em}\\
    \includegraphics[width=8.2cm]{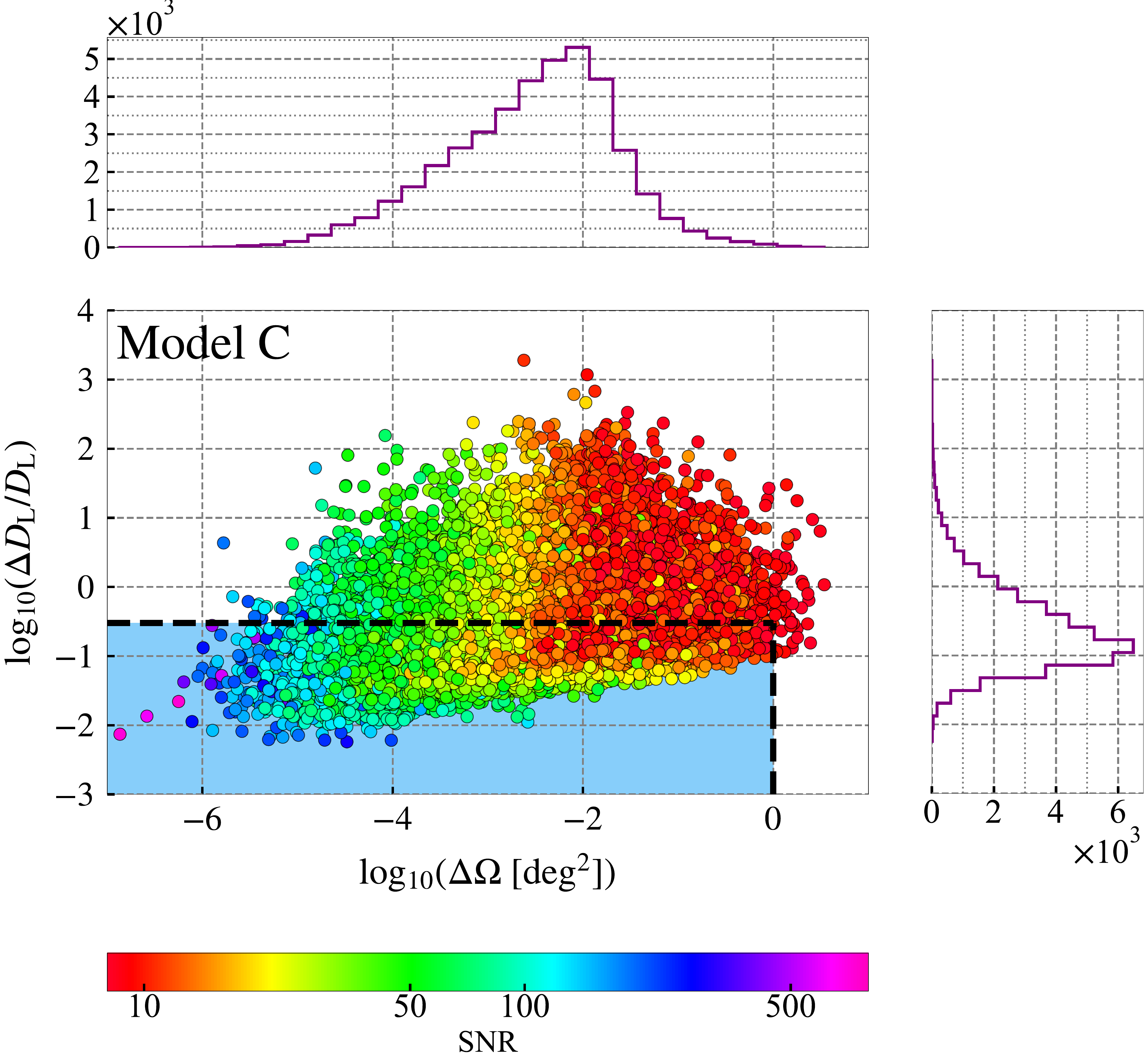}
    \hspace{2.5em}
    \includegraphics[width=8.2cm]{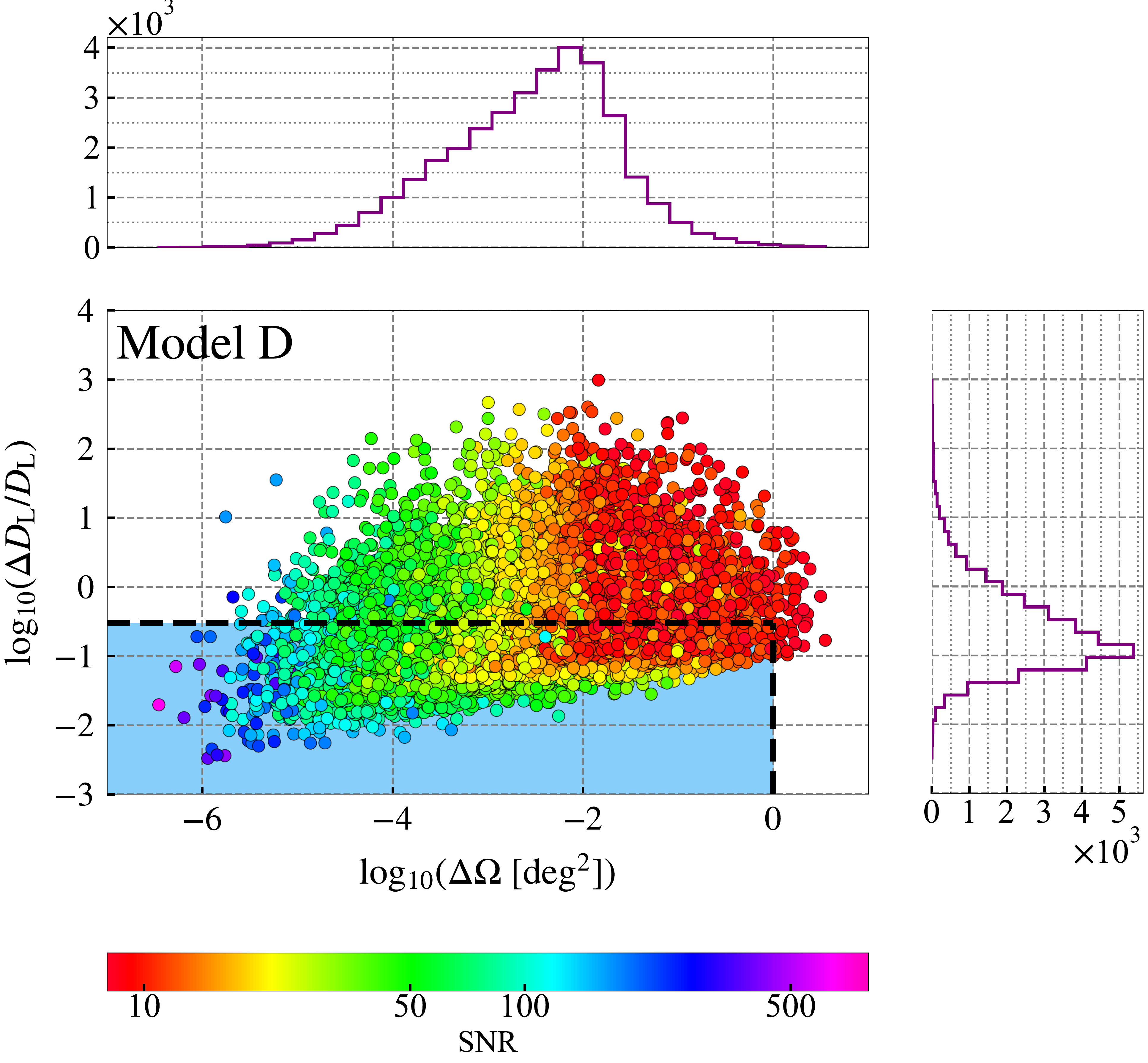}
    \caption{Same as Figure~\ref{ fig:DL_DOmega_DO }, but for DEC. Note that all
    detectable NS-WD merger events are color-coded by the logarithmic SNRs in
    each plot.} 
    \label{ fig:DL_DOmega_DEC }
\end{figure*}

Furthermore, based on the GW early-warning NS-WD mergers, we now present
estimations for the accuracy of distance ($\Delta D_{\mathrm{L}}$) and angular
resolution ($\Delta \Omega$), which will be of great use to synergy observations
with EM facilities. As described in Section~\ref{ sec:warnings }, our analyses
are performed using the FIM method. We first compare ${\Delta
D_{\mathrm{L}}}/{D_{\mathrm{L}}}$-$\Delta \Omega$ distributions of our yearly GW
early-warning samples with DO-OPT for the four population models in Figure~\ref{
fig:DL_DOmega_DO }; similar distributions for DEC are plotted in Figure~\ref{
fig:DL_DOmega_DEC }. We mark all detectable NS-WD mergers with circles in
different colours based on their SNRs. Our results seem to suggest that the
mergers with higher SNRs tend to have better localization accuracies (i.e.,
smaller $\Delta \Omega$ and $\Delta D_{\mathrm{L}}/D_{\mathrm{L}}$), which are
mainly distributed in the lower left of these plots in Figure~\ref{
fig:DL_DOmega_DO } and Figure~\ref{ fig:DL_DOmega_DEC }. In each plot, the
purple histograms show the $\Delta \Omega$ and $\Delta
D_{\mathrm{L}}/D_{\mathrm{L}}$ distributions in upper and right panels,
respectively. Both Figure~\ref{ fig:DL_DOmega_DO } and Figure~\ref{
fig:DL_DOmega_DEC } show that the peak of the $\Delta \Omega$ distribution is at
$\simeq \mathcal{O}(0.01)\,\mathrm{deg}^2$ level, regardless of the population
models in our considerations. For the $\Delta D_{\mathrm{L}}/D_{\mathrm{L}}$
distributions, most early-warning samples of DO-OPT are clustered in the range
of $0.3 \lesssim \Delta D_{\mathrm{L}}/D_{\mathrm{L}} \lesssim 1$. In
comparison, DEC provides more accurate estimations of $\Delta D_{\mathrm{L}}$,
given that the peak of the $\Delta D_{\mathrm{L}}/D_{\mathrm{L}}$ distribution
is mostly at $\simeq \mathcal{O}(0.1)$ level.  

To go a step further, we are particularly interested in those mergers that would
yield the best estimation results of the distance and angular resolution. This
is because these events can provide helpful inputs for future multimessenger
astronomy with decihertz GW early warnings and EM follow-ups. In view of this,
we further define a Golden Sample (denoted as `GS' hereafter) set for our yearly
GW early-warning samples with the two criteria that
\begin{enumerate}[(i)] 
  \item $\Delta \Omega < 1\,\mathrm{deg}^2$, and
  \item $\Delta D_{\mathrm{L}}/D_{\mathrm{L}} < 0.3$.
\end{enumerate}
Given that many current and planned wide-field optical survey projects have FoV
$\gtrsim 1\,\mathrm{deg}^2$ (e.g., see in Table~\ref{ tab:Survey projects } the
summary of the technical information for some optical survey telescopes), we
follow the discussions in \citet{Kang:2022nmz} to set $1\,\mathrm{deg}^2$ to be
the threshold in criterion~(i). For criterion~(ii), we arbitrarily set 30\% to
be the selection criterion on $\Delta D_{\mathrm{L}}/D_{\mathrm{L}}$. Note that
similar crude treatments in fact are commonly adopted in many studies
\citep[e.g., ][]{Tamanini:2018cqb, Kang:2021bmp}. Based on the
above analyses, the NS-WD mergers in GS would matter the most to future
multimessenger early-warning detections. In Figure~\ref{
fig:DL_DOmega_DO } and Figure~\ref{ fig:DL_DOmega_DEC } we plot black dashed
lines to denote the selection functions corresponding to the criteria~(i) and
(ii) above. Therefore, the shaded blue regions in Figure~\ref{ fig:DL_DOmega_DO
} and Figure~\ref{ fig:DL_DOmega_DEC } delimit the parameter space of GS. 

We list in Table~\ref{ tab:GS } yearly detection numbers of NS-WD mergers in GS
for the four population models with DO-OPT and DEC. The percentages in brackets
show the ratio to the total number of yearly detectable NS-WD mergers (see
Table~\ref{ tab:detection rates }). Combining all the results shown in
Table~\ref{ tab:GS }, we conclude that DEC performs much better on GS detections
as a whole, with its yearly detection number in the range of $\sim (2.1
\mbox{--} 2.8) \times 10^{4}$. In comparison, the GS detection number for DO-OPT
is in the range of $\sim (3.8 \mbox{--} 4.5) \times 10^{2}$, which is
approximately 50 times fewer than DEC. In particular, we note that the
percentage of GS detections with DEC to its total yearly detectable mergers is
$\gtrsim 60\%$, while it is much smaller for DO-OPT, with the ratio to be
$\lesssim 25\%$. 

Finally, let us move to the potential for future multimessenger early-warning
detections of NS-WD mergers. Taking the peculiar lGRB, GRB\,211211A, as an
example,\footnote{Recently, there is another extremely bright cb-lGRB,
GRB\,230307A, at a redshift of $z=0.065$ \citep[$\approx
300\,\mathrm{Mpc}$;][]{Fausnaugh:2023pmx, JWST:2023jqa, Sun:2023rbr,
Mereghetti:2023lar, Gillanders:2023zys}. However, given the different features
of the afterglow/kilonova-like emissions and precursor properties from
GRB\,211211A, most investigation was not in favor of the NS-WD merger origin for
GRB\,230307A \citep{Wang:2023yky, JWST:2023jqa, Dichiara:2023goh, Song:2023tdb,
Yang:2023mqt}. It is more likely to be a BNS merger.} the observational facts
show that it is a nearby bright GRB at $z=0.0763$ ($\approx 350\,\mathrm{Mpc}$)
with a possible NS-WD merger origin \citep{Rastinejad:2022zbg, Troja:2022yya,
Yang:2022qmy, Mei:2022ncd, Gompertz:2022jsg}. Using a set of pseudo-GRBs,
\citet{Yin:2023gwc} have shown that a GRB\,211211A-like event could be
detectable up to a maximum redshift $z=0.52$ ($\approx 3000\,\mathrm{Mpc}$). On
the other hand, our results suggest that the number of yearly detectable NS-WD
mergers with $z \lesssim 0.52$ is in the range of $\sim 990 \mbox{--} 1300$ for
DO-OPT, and the corresponding percentage to the total number of yearly
detectable merger events (see Table~\ref{ tab:detection rates }) is $\sim 65
\%$. They are $\sim 5200 \mbox{--} 6700$ and $\sim 15 \%$ for DEC. \add{Remember that GRBs are highly beamed, so another condition for the detection of a successful GRB event is that its jet should beam towards the Earth.\footnote{\add{In the case of a choked jet, the jet energy would be stored in a hot,} \add{shocked material, forming a broad cocoon without a spine jet \citep{Zhang:2018ads}. Choked jets could have larger viewing angles and be even more abundant than those that successfully break out. This topic deserves more detailed analyses.}} The beaming factor of a GRB is defined} \add{by ${f_{\mathrm{b}} \equiv \Delta \Omega_{\mathrm{b}}/4 \pi \simeq \theta_{\mathrm{v}}^{2}/2}$, where $\Delta \Omega_{\mathrm{b}}$ is the solid angle of a bipolar, conical jet, and $\theta_{\mathrm{v}}$ ($\ll 1$) is its half-opening angle. Assuming a typical value of $15^\circ$ for $\theta_{\mathrm{v}}$, we could obtain ${f_{\mathrm{b}} \simeq 0.034}$. Considering such relativistic beaming effects, there should be a few tens (hundreds) yearly detectable NS-WD mergers that beam towards us for DO-OPT (DEC) within $z \lesssim 0.52$. Given that the properties of NS-WD GRBs are still under debate \citep[see, e.g.,][]{King:2006nw, Zenati:2018gcp, Bobrick:2021hho, Moran-Fraile:2023oui, Ablimit:2023srq}, we point out here that if there are $\gtrsim 1$\% of NS–WD mergers with $z \lesssim 0.52$ that can successfully produce GRB\,211211A-like signals, the possibility of multimessenger early-warning detections indeed exists during a 4-yr mission time for decihertz GW detectors. This also means that future multimessenger early-warning detections could help people constrain the fraction of NS-WD systems that are capable of producing GRBs.} 

\begin{table*}
    \renewcommand\arraystretch{2}
    \centering
    \caption{Yearly detection numbers and percentages (in brackets) of NS-WD
    mergers in the GS for different population models with DO-OPT and DEC. Note
    that the percentages in brackets show the ratio to the total number of
    yearly detectable NS-WD mergers (see Table~\ref{ tab:detection rates }).}
    \setlength{\tabcolsep}{1.1cm}{\begin{tabular}{c c c c c}
    \toprule
    \toprule
    \vspace{-3.5em}\\
    GW Detector & \multicolumn{4}{c}{Population Model}\\  
    \cline{2-5}
    & A    & B    & C    & D\\
    
    \cline{1-5}
    
    \multirow{2}{30pt}{\shortstack{DO-OPT}} 
    & 451    
    & 396    
    & 428    
    & 378\vspace{-1em}\\  
            
    & (23.55\,\%)            
    & (24.19\,\%)          
    & (24.85\,\%)            
    & (25.39\,\%)\\
    
    \cline{1-5}
    
    \multirow{2}{18pt}{\shortstack{DEC}}
    & 28229    
    & 22320    
    & 25301    
    & 20677\vspace{-1em}\\   
            
    & (61.30\,\%)           
    & (61.77\,\%)  
    & (61.30\,\%)         
    & (62.07\,\%)\vspace{0em}\\
    \bottomrule
    \end{tabular}}
    \label{ tab:GS }
\end{table*}

Besides the peculiar GRB signal, optical/infrared kilonova emissions associated
with GRB\,211211A were also observed. Many recent studies have proposed various
kilonova models and suggest that the peak AB absolute magnitude of
GRB\,211211A-like kilonovae should be $\approx {-17}$ in $\it{g}$, $\it{r}$, and
$\it{i}$ bands \citep{Yang:2022qmy, Barnes:2023ixp, Kunert:2023vqd,
Zhong:2023zwh}. Based on the above finding, if we set $z=0.52$ as the redshift
upper  limit for  detections of GRB\,211211A-like events, the limiting magnitude
$m^*$ for wide-field optical survey projects should satisfy $m^* \gtrsim 25$ to
ensure the kilonova detections. From Table~\ref{ tab:Survey projects } we find
that LSST and CSST meet the above requirements well. Therefore, more cooperative
observations with different optical surveys and decihertz GW detectors would be
very promising in the future. Given that the true nature of GRB\,211211A-like
events remains unknown, we emphasize that the multimessenger observations with
GW early warnings and the above multiband EM follow-ups would shed light on the
origin of such peculiar lGRBs. Note that even for the NS-WD merger scenario, the
properties of their EM follow-up transients could vary widely, depending on the
component masses, the magnetic field strength and configuration, specific
afterglow and kilonova models, {\it etc.}.\footnote{For example,
\citet{Yang:2022qmy} pointed out that a massive WD component near the
Chandrasekhar mass limit should be involved to trigger the accretion-induced
collapse of the WD during the GRB\,211211A event.} We leave the impacts of those
uncertainties on the realistic multimessenger observations for future studies.

\section{Conclusion}
\label{ sec:Conclusion }

In this work, we follow \citet{Liu:2022mcd} to discuss GW detections and
early-warning predictions of NS-WD mergers with two space-borne decihertz GW
detectors, DO-OPT and DEC. Based on different SFR and DTD models, we use the
method of convolution with a series of Monte Carlo simulations to obtain four
NS-WD merger population models. For all NS-WD mergers with the merger time
$1\,\mathrm{yr} \leq t_{\mathrm{c}_{0}} \leq 2\,\mathrm{yr}$, we set the
early-warning time to $t_\mathrm{e} = 1\,\mathrm{d}$ and SNR threshold to 8 to
perform  GW detection strategy. Based on the NS-WD merger populations, we not
only give quick assessments of GW detection rates with the two decihertz GW
detectors, but also report systematic analyses on the characteristics of
GW-detectable merger events using the FIM. We find that DEC has better
performance than DO-OPT as a whole, especially for high-$z$ events. The yearly
GW detection number for DEC is in the range of $\sim (3.3 \mbox{--} 4.6) \times
10^{4}$, while it is only $\sim (1.5 \mbox{--} 1.9) \times 10^{3}$ for DO-OPT,
approximately 20 times fewer. Taking the population Model A as an example,
DO-OPT can detect early-warning sources up to $z \lesssim 1.7$, while DEC could
reach $z \lesssim 7.6$. Moreover, detectable NS-WD mergers are clustered at $z
\simeq 0.8$ for DEC. In comparison, due to the limited detection abilities, most
detectable NS-WD mergers are clustered around $z \simeq 0.25$ for DO-OPT.
Similarly, for the maximum SNR, we show that DEC reaches $\mathrm{SNR} \simeq
650$, while DO-OPT  reaches $\mathrm{SNR} \simeq 75$.

Differently from BNS mergers or NS-BH mergers, we note that GW early-warning
detections of NS-WD mergers are essentially to offer alerts on the RLOF time,
especially considering that there is no well-modeled GW waveform for NS-WD
systems during the runaway mass-transfer phase. In view of this, for the FIM in
the frequency domain, the integration upper limit $f_{\text {out}}$ should be
related to the  GW cutoff frequency at the critical RLOF separation. We plot the
distributions of $f_{\text {out}}$ for detectable NS-WD mergers observed by
DO-OPT and DEC. The majority of total detectable NS-WD mergers for DEC cluster
at $f_{\text {out}} \simeq 0.1\,\mathrm{Hz}$, while in comparison, the peak of
$f_{\text {out}}$ distribution for DO-OPT is a little higher in a wider range
($0.16 \mbox{--} 0.25\,\mathrm{Hz}$). These could be explained by the
cosmological redshift effect, given that DEC will detect more high-$z$ events. 

As for the estimations of the accuracy of distance and angular resolution, we
compare ${\Delta D_{\mathrm{L}}}/{D_{\mathrm{L}}}$-$\Delta \Omega$ distributions
of yearly GW early-warning samples for DO-OPT and DEC with different population
models. Our results show that the peak of the $\Delta \Omega$ distribution is
expected to be at $\simeq \mathcal{O}(0.01)\,\mathrm{deg}^2$ level, regardless
of the changes in decihertz GW detectors and population models. For the $\Delta
D_{\mathrm{L}}/D_{\mathrm{L}}$ distributions, most early-warning samples of
DO-OPT are clustered in the range of $0.3 \lesssim \Delta
D_{\mathrm{L}}/D_{\mathrm{L}} \lesssim 1$, while in comparison, DEC can provide
more accurate estimations of $\Delta D_{\mathrm{L}}$, given that the peak of the
$\Delta D_{\mathrm{L}}/D_{\mathrm{L}}$ distribution is at $\simeq
\mathcal{O}(0.1)$ level. Furthermore, for those mergers that would yield the
best estimation results of the distance and angular resolution, we define them
as a GS. We present the yearly detection numbers of NS-WD mergers in GS for
different population models with DO-OPT and DEC. We conclude that DEC performs
much better on GS detections as a whole, with its yearly detection number in the
range of $\sim (2.1 \mbox{--} 2.8) \times 10^{4}$, while it is $\sim (3.8
\mbox{--} 4.5) \times 10^{2}$ for DO-OPT, approximately 50 times fewer than DEC.
The percentage of GS detections with DEC to its total yearly detectable NS-WD
mergers could achieve $\gtrsim 60\%$, while it is much smaller for DO-OPT with
the ratio to be  $\lesssim 25\%$. 

Finally, taking the recent peculiar lGRB, GRB\,211211A, as an example, we
further discuss the potential for future multimessenger early-warning detections
of NS-WD mergers. With a sufficient early-warning time (e.g., $t_\mathrm{e} =
1\,\mathrm{d}$) and the localization accuracies, we suggest that the GW
early-warning detection will allow future EM telescopes to prepare well in
advance for the possible follow-up transients after some special NS-WD mergers.
Given that the nature of cb-lGRBs remains an open question, the multimessenger
observations with the GW early warnings and the multiband EM follow-ups are
expected to shed new light on their properties. For example, compared with BNS
and NS-BH scenarios, no inspiral GW signals of NS-WD mergers are expected to
enter the ground-based detectors following decihertz GW early warnings. The EM
follow-up transients with different origins may also vary widely. In view of
this, our results can provide meaningful references and helpful inputs for
upcoming EM follow-up projects.

There are, of course, many ways in which this study can be extended. At the end
of Section~\ref{ sec:warnings }, we have emphasized that precise GW-waveform
constructions should take into account the effects of mass transfer and
dynamical tides for WDs, especially in the late inspiral stages of NS-WD
mergers. \add{Also, our results leave out the consideration of the confusion noise \citep{Christensen:2018iqi}, which depends strongly on the specific population model, the sensitivity of the detector, and its operation time, {\it etc.}.\footnote{\add{It has been shown that the confusion noise from DCOs could be} \add{subtracted out by an iteration scheme and a global fit \citep{Cutler:2005qq}. More related discussions on the impact of confusion noise for decihertz GW detectors can be found in \citet{Liu:2022mcd}.}} We also leave out the dedicated analyses of overlapping signals which may affect the parameter estimation \citep[see e.g., ][]{Wang:2023ldq}, but we expect our result unlikely to 
change significantly even when overlapping signals are considered. Moreover, the} \add{realistic multimessenger searching strategy, including the communication between GW detectors and EM telescopes, should be considered.} On the other hand, even if there are no observable EM transients after NS-WD mergers, we suggest that the GW-detectable NS-WD populations in GS could still be of great use for the dark-siren cosmology \citep[see, e.g.,][]{Schutz:1986gp, LIGOScientific:2018gmd, DES:2019ccw, Zhu:2021aat, Zhu:2021bpp, Liu:2022rvk, Yang:2022fgp, Seymour:2022teq}. We hope that more studies can be carried out in the future.

\section*{Acknowledgements}

\add{We would like to thank the anonymous referee for helpful and valuable comments and suggestions.} This work was supported by 
the Beijing Municipal Natural Science Foundation (1242018),
the National Natural Science Foundation of China
(11975027, 11991053, 11721303, U2038105, U1831135, 12121003, 12103065), the National SKA
Program of China (2020SKA0120300, 2022SKA0130102), the National Key Research and
Development Programs of China (2022YFF0711404), the science research grants from
the China Manned Space Project with No.\ CMS-CSST-2021-B11, the Max Planck
Partner Group Program funded by the Max Planck Society, the Fundamental Research
Funds for the Central Universities,  the Program for Innovative Talents and
Entrepreneur in Jiangsu, and the High-Performance Computing Platform of Peking
University. C.L.\ was supported by the China Scholarship Council (CSC).

\section*{Data Availability}

The data underlying this paper will be shared on a reasonable request to the
corresponding authors.



\bibliographystyle{mnras}
\bibliography{refs} 





\appendix

\section{The redshift distribution factor}
\label{ sec:appA }

As mentioned in Section~\ref{ sec:populations }, when we ignore the possible
redshift evolution of intrinsic system parameters for NS-WD mergers, the
redshift-dependent merger event rate density $\dot{\rho}(z)$ is, 
\begin{equation}
\dot{\rho}(z)=\dot{\rho}_{0} f(z) \,,
\label{ eq:dotrho }
\end{equation}
where $z$ is the redshift, $\dot{\rho}_{0}$ is the local NS-WD event rate
density, and $f(z)$ is the dimensionless redshift distribution factor. On the
other hand, $\dot{\rho}(z)$ can also connect to the cosmological SFR density
$\dot{\rho}_{*}(z)$ by accounting for the probability density function of delay
time $P(\tau)$ of NS-WD mergers via 
\begin{equation}
\begin{aligned} \dot{\rho}(z) & \propto \int_{\tau_{\min }}^{\tau_{\max }}
\dot{\rho}_{*}\left[z^{\prime}(\tau)\right] P(\tau) d \tau \\ 
    & \propto \int_{z^{\prime}_{\min}}^{z^{\prime}_{\max}}
    \dot{\rho}_{*}\left(z^{\prime}\right)
    P\left[\tau\left(z^{\prime}\right)\right] \left[-\frac{d
    t\left(z^{\prime}\right)}{d z^{\prime}}\right] d z^{\prime} \,,\end{aligned}
\label{ eq:dotrho DTD }
\end{equation}
where $\tau=t(z)-t\left(z^{\prime}\right)$ is the delay time\footnote{As defined
in Section~3.2.1 of \citet{Toonen:2018njy}, the delay time is the evolutionary
time between the merger and formation of the binary with two zero-age main
sequence stars.} and $t\left(z^{\prime}\right)$ is the time when the binaries
are formed with two zero-age main sequence stars, $t(z)$ is the merger time for
such binaries, and $\tau_{\min }$ and $\tau_{\max}$ are the minimum and maximum
delay times, respectively. 

With Equations~(\ref{ eq:dotrho }) and (\ref{ eq:dotrho DTD }), we can obtain
$f(z)$ by normalizing it to unity in the local Universe (i.e., $z = 0$). Note
that here $t$ is the age of the Universe and we have, 
\begin{equation}
\frac{d t(z)}{d z}=-\frac{1}{H_{0}(1+z) \sqrt{\Omega_{\Lambda} + (1+z)^{3}
\Omega_{\mathrm{m}}}} \,.
\label{ eq:dtdz }
\end{equation}
Therefore, for NS-WD mergers at a fixed $z$, we can calculate the corresponding
$z^{\prime}_{\min}$ by the definition of $\tau_{\min }$ and Equation~(\ref{
eq:dtdz }). Based on the population-synthesis results presented in
\citet{Toonen:2018njy}, we set $\tau_{\min} = 0.1\,\mathrm{Gyr}$ in our
calculations. Following \citet{Khokhriakova:2019uzo}, we also set the upper
integration redshift limit $z^{\prime}_{\max}$ to be 11.247 (corresponding to
the time when the age of our Universe was $400\,\mathrm{Myr}$) when the
intensive star formation began.

For the SFR density $\dot{\rho}_{*}(z)$, we consider two kinds of analytical
models from \citet{Yuksel:2008cu} and \citet{Madau:2014bja} (abbreviated as
`Y08' and `MD14', respectively) in our work. They read,
\begin{align}
\dot{\rho}_{*}^{\mathrm{Y08}}(z) &\propto {\left[(1+z)^{3.4
\eta}+\left(\frac{1+z}{5000}\right)^{-0.3 \eta}\right.}
\left.+\left(\frac{1+z}{9}\right)^{-3.5 \eta}\right]^{1 / \eta} \,, \label{
eq:SFR08 } \\
\dot{\rho}_{*}^{\mathrm{MD14}}(z) &\propto \frac{(1+z)^{2.7}}{1+[(1+z) /
2.9]^{5.6}} \,.  \label{ eq:SFR14 }
\end{align}
Note that $\eta=-10$ is adopted in Equation~(\ref{ eq:SFR08 }). 

\begin{table}
    \renewcommand\arraystretch{1.3}
    \centering
    \caption{The best-fit $\alpha_{\mathrm{t}}$ values derived from different
    DTDs shown in Section~3.2.1 of \citet{Toonen:2018njy} by Equation~(\ref{
    eq:PL model }). We consider four combinations of two CE evolution models
    (model `$\alpha \alpha$' and model `$\gamma \alpha$') and two SN-kick
    velocity distributions (abbreviated as `$\mathrm{H}$' and `$\mathrm{V}$').
    More details about these models are given in Appendix~\ref{ sec:appA }.}
    \begin{threeparttable}
    \setlength{\tabcolsep}{16.0mm}{\begin{tabular}{cc}
    \toprule
    \toprule
    Model    & $\alpha_{\mathrm{t}}$ \\
    \midrule 
    $\alpha \alpha$-$\mathrm{H}$        & $-1.12$\\
    $\alpha \alpha$-$\mathrm{V}$        & $-0.98$\\
    $\gamma \alpha$-$\mathrm{H}$        & $-1.11$\\
    $\gamma \alpha$-$\mathrm{V}$        & $-0.95$\vspace{0.3em}\\ 
    \bottomrule
    \end{tabular}}
    \end{threeparttable}
    \label{ tab:PL fitting }
\end{table}

\begin{figure*}
    \centering
    \includegraphics[width=8.2cm]{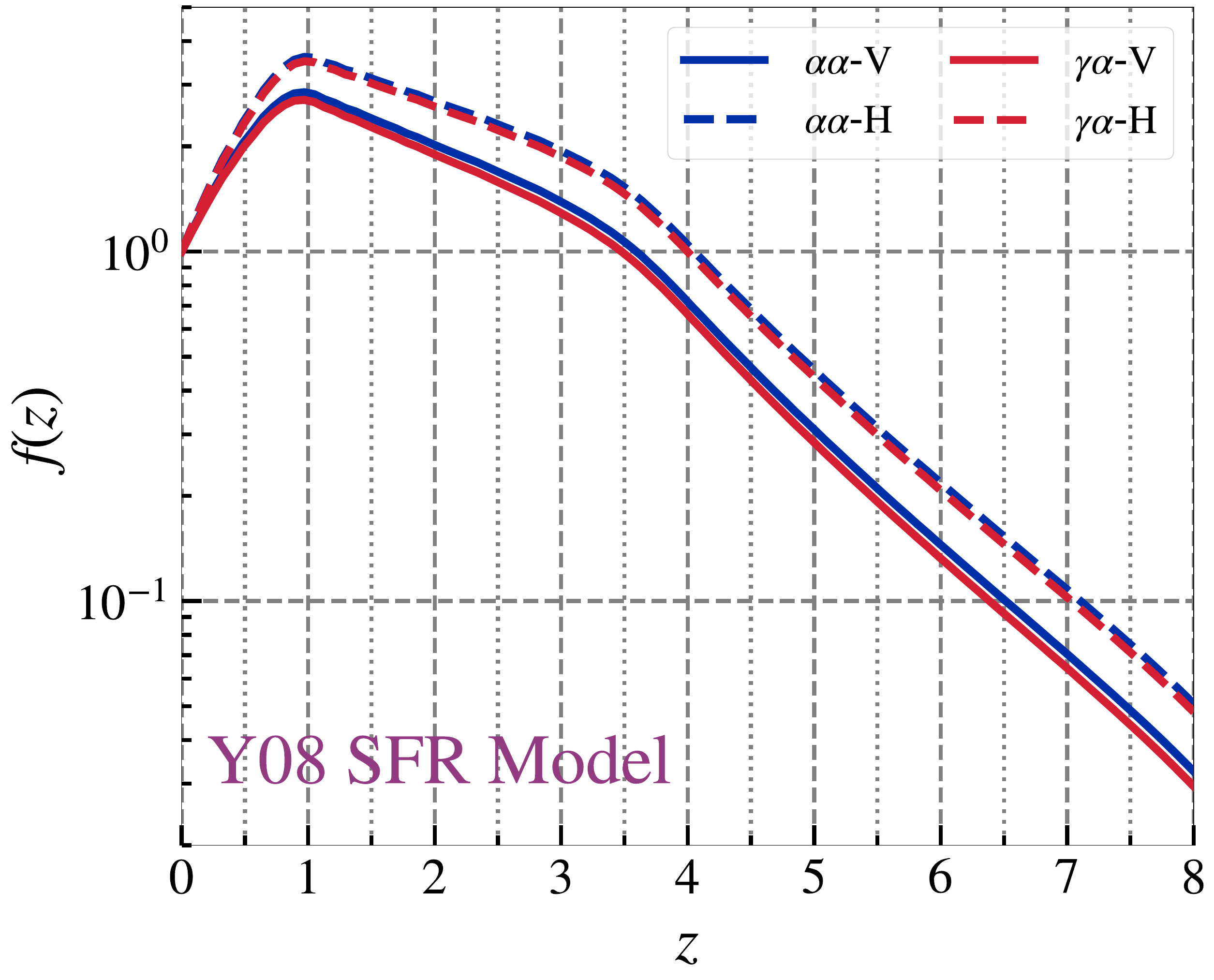}
    \hspace{2.5em}
    \includegraphics[width=8.2cm]{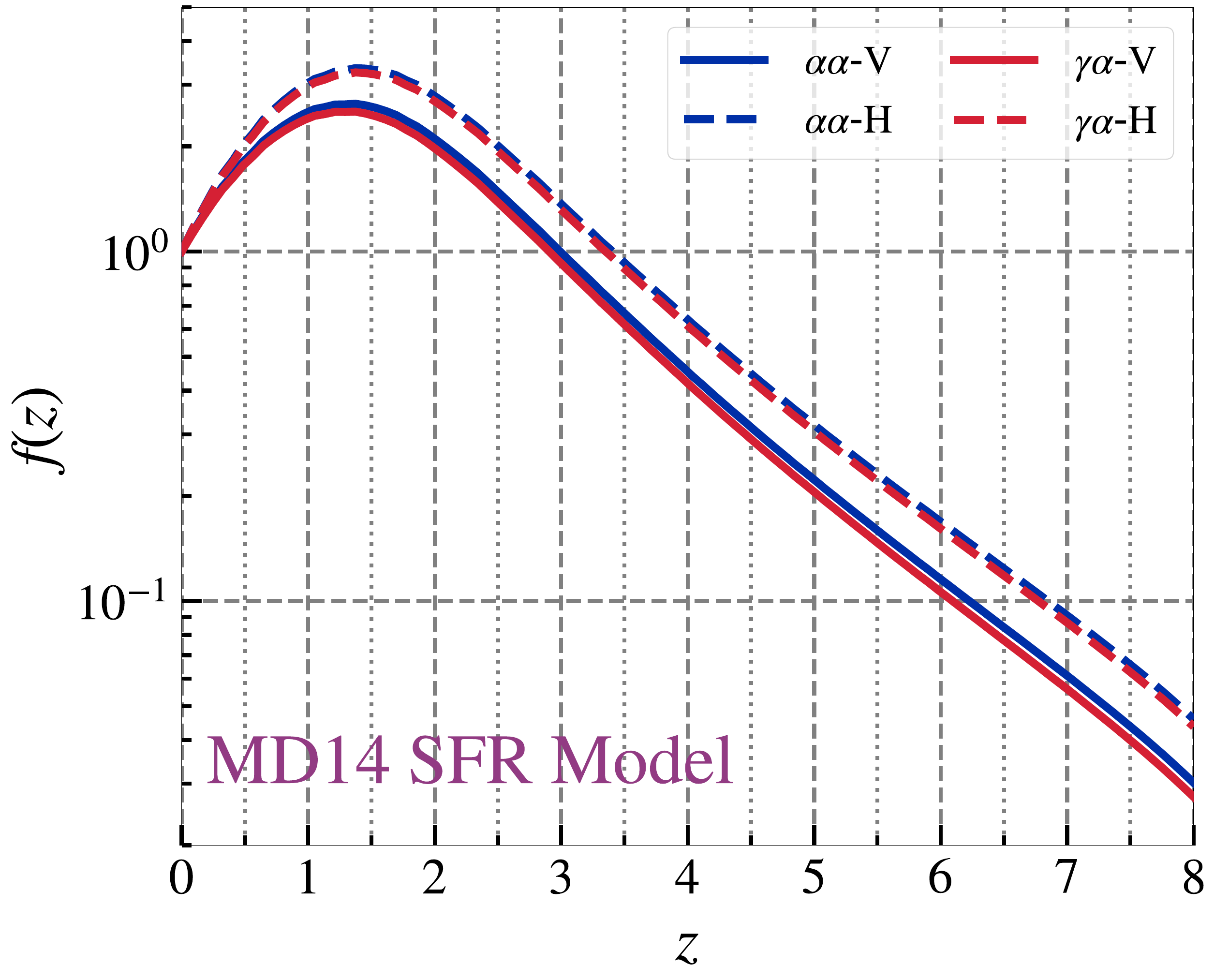}
    \caption{Redshift distributions of $f(z)$ derived from the method of
    convolution by Equation~(\ref{ eq:dotrho DTD }). The Y08 SFR model is
    adopted in the left plot, while the MD14 model is applied in the right plot.
    The coloured solid lines and dashed lines denote different DTD models. The
    redshift cutoff  $z = 8$ is adopted here.} 
    \label{ fig:fz }
\end{figure*}

\begin{table*}
    \renewcommand\arraystretch{1.6}
    \centering
    \caption{Technical information for some wide-field optical survey projects
    from \citet{Zhu:2020ffa, Zhu:2021ram}. Note that a 300-s exposure time is
    adopted for the search limiting magnitude $m^*$ in the three most common
    filters ($\it{g}, \it{r}, \it{i}$). The references are: (1)
    \citet{2019PASP..131a8003M}, \citet{2019PASP..131a8002B}; (2)
    \citet{LSSTScience:2009jmu}, \citet{LSST:2008ijt}; (3)
    \citet{2018AcASn..59...22S}, \citet{WFST:2023voz}; (4)
    \citet{2020SPIE11445E..7MY}; (5) \citet{Gong:2019yxt}.} 
    \setlength{\tabcolsep}{0.71cm}{\begin{tabular}{c ccc c c c}
    \toprule
    \toprule
    \vspace{-2.5em}\\
    Survey        & \multicolumn{3}{c}{$m^*$/$\mathrm{mag}$}
                     & FoV/$\mathrm{deg}^2$  
                     & Sky Coverage/$\mathrm{deg}^2$  
                     & Reference\vspace{0.4em}\\
    \cline{2-4}
                     & $\it{g}$ & $\it{r}$ & $\it{i}$\\
    
    \midrule
    ZTF        & 21.6        & 21.3        & 20.9        & 47.7         
               & 30000       & (1)\\
    LSST       & 26.2        & 25.7        & 25.8        & 9.6        
               & 20000       & (2) \\
    WFST       & 24.2        & 24.0        & 23.3        & 6.55         
               & 20000       & (3)\\
    Mephisto   & 24.2        & 23.9        & 23.4        & 3.14         
               & 26000       & (4)\\
    CSST       & 26.3        & 26.0         & 25.9         & 1.1         
               & 17500       & (5)\vspace{0.3em}\\ 
    \bottomrule
    \end{tabular}}
    \label{ tab:Survey projects }
\end{table*}

As for the probability density function of delay time $P(\tau)$, many studies
have assumed some empirical forms and derived the best-fit parameters with
simulated data. Such a treatment has been widely applied to other DCO systems
\citep[e.g., BNS and NS-BH systems; ][]{Virgili:2009ca, Wanderman:2014eza,
Sun:2015bda, Zhu:2020ffa}. Therefore, for simplicity, we assume a power-law
dependence for the DTD of NS-WD mergers, 
\begin{equation}
P(\tau) \propto \tau^{-\alpha_{\mathrm{t}}} \,,
\label{ eq:PL model }
\end{equation}
where $\alpha_{\mathrm{t}}$ is a phenomenological parameter. Considering
different common-envelope (CE) evolutions\footnote{As a mass-loss phase in the
formation of DCO systems, CE evolution can lead to a severe shrinkage of the
binary orbit. We refer readers to see \citet{Ivanova:2012vx} for a comprehensive
review.} and different NS natal-kick\footnote{NS natal-kicks can also be called
SN-kicks. Many studies of pulsar scale heights \citep{1970ApJ...160..979G},
proper motions of pulsars \citep{1993Natur.362..133C, Lyne:1994az, Hobbs:2005yx,
Verbunt:2017zqi}, and high velocities of some single NSs
\citep{Chatterjee:2005mj, Becker:2012dj} have suggested that a kick should be
imparted to the NS during core-collapse SNs.} distributions,
\citet{Toonen:2018njy} have recently presented different DTDs of NS-WD mergers.
Here, we fit four kinds of DTDs taken from \citet{Toonen:2018njy}---abbreviated
as `$\alpha \alpha$-$\mathrm{H}$', `$\alpha \alpha$-$\mathrm{V}$', `$\gamma
\alpha$-$\mathrm{H}$', and `$\gamma \alpha$-$\mathrm{V}$'---by Equation~(\ref{
eq:PL model }) and derive the best-fit $\alpha_{\mathrm{t}}$ values, which are
listed in Table~\ref{ tab:PL fitting }. Note that `$\alpha$' and `$\gamma$' are
commonly used to denote different treatments for the CE phase where the
`$\alpha$'-formalism is based on the energy conservation while the
`$\gamma$'-formalism is based on the balance of angular momentum instead of the
energy \citep{Ivanova:2012vx}. More specifically, following
\citet{Toonen:2018njy}, we use `$\alpha \alpha$' and `$\gamma \alpha$' to denote
two kinds of CE evolutionary models for NS-WD mergers where the
`$\alpha$'-prescription is used to determine the outcome of every CE in model
`$\alpha \alpha$'. In model `$\gamma \alpha$', the `$\gamma$'-prescription is
introduced to describe the first CE phase, while the `$\alpha$'-prescription is
applied in the second CE phase. For SN-kick velocity distributions, we adopt two
kinds of models from \citet{Hobbs:2005yx} and \citet{Verbunt:2017zqi},
respectively (abbreviated as `$\mathrm{H}$' and `$\mathrm{V}$' when combined
with model `$\alpha \alpha$' and model `$\gamma \alpha$'). The former (i.e.,
`$\mathrm{H}$') is a Maxwellian distribution with a one-dimensional root mean
square of $\sigma_{\mathrm{H}} = 265\,\mathrm{km}\,\mathrm{s}^{-1}$, while the
latter (i.e., `$\mathrm{V}$') includes two Maxwellian components with
$\sigma_{\mathrm{V}}^{1} = 75\,\mathrm{km}\,\mathrm{s}^{-1}$ and
$\sigma_{\mathrm{V}}^{2} = 316\,\mathrm{km}\,\mathrm{s}^{-1}$ from a direct
comparison of pulsar parallaxes and proper motions. More detailed descriptions
of the above DTDs can be found in \citet{Toonen:2018njy}.

Based on the above SFR and DTD models, we follow \citet{Zhu:2020ffa} to use the
method of convolution by Equation~(\ref{ eq:dotrho DTD })  and obtain the
dimensionless redshift distribution factor $f(z)$, which are shown in
Figure~\ref{ fig:fz }. Given a specific DTD model, Figure~\ref{ fig:fz } shows
that $f(z)$ has a dramatic dependence on the SFR model in the relatively lower
redshift regime ($z \lesssim 4.5$), while in the high-$z$ regime ($z > 4.5$),
there is little difference between the Y08 SFR model and the MD14 model. On the
other hand, when we adopt the same SFR model, both Table~\ref{ tab:PL fitting }
and Figure~\ref{ fig:fz } suggest that different CE evolution models will not
influence $f(z)$ significantly, while $f(z)$ can be indeed influenced by the
choice of the SN-kick velocity distribution. For this reason, we decide to
report detailed analyses only between the model $\gamma \alpha$-$\mathrm{H}$ and
the model $\gamma \alpha$-$\mathrm{V}$ in the main text for more in-depth
comparisons. 

\section{Information for optical surveys}
\label{ sec:appB }

Technical parameters for five current and planned optical time-domain surveys
are listed in Table~\ref{ tab:Survey projects }, which are: the Zwicky Transient
Facility \citep[ZTF;][]{2019PASP..131a8003M, 2019PASP..131a8002B}; the Large
Synoptic Survey Telescope \citep[LSST, newly named as the Vera Rubin
Observatory;][]{LSSTScience:2009jmu, LSST:2008ijt}; the Wide Field Survey
Telescope \citep[WFST;][]{2018AcASn..59...22S, WFST:2023voz}; the Multi-channel
Photometric Survey Telescope \citep[Mephisto;][]{2020SPIE11445E..7MY}; and the
Chinese Space Station Telescope \citep[CSST;][]{Gong:2019yxt}. Given the search
limiting magnitude $m^*$, one can obtain the effective limiting flux $F_{\nu}^*$
for each survey telescope by $F_{\nu}^* \simeq 3631\,\mathrm{Jy} \times 10^{-m^*
/ 2.5}$. Note that Table~\ref{ tab:Survey projects } only lists $m^*$ values
with the three most common bands ($\it{g}, \it{r}, \it{i}$), assuming a 300-s
exposure time for each survey. 

Recently, \citet{Kang:2022nmz} have investigated the prospects of multimessenger
early-warning detections for BNS mergers, with the help of early warnings from
different decihertz GW observatories and optical survey missions. We leave
similar analyses on NS-WD mergers for future studies.


\bsp	
\label{lastpage}
\end{document}